\PassOptionsToPackage{table}{xcolor} 
\documentclass[compsoc,conference,a4paper,10pt,times]{IEEEtran}
\IEEEoverridecommandlockouts
\usepackage{cite}
\usepackage{url}
\usepackage{xcolor}
\usepackage[colorlinks=true,urlcolor=black,linkcolor=black,citecolor=black]{hyperref}
\usepackage{microtype}
\usepackage[most]{tcolorbox}

\usepackage{tikz}
\usepackage{amsmath}
\usetikzlibrary{arrows.meta, positioning, shapes.geometric, fit, calc, backgrounds, decorations.pathreplacing}
\usepackage{adjustbox} 
\usepackage{graphicx}
\usepackage{enumitem}

\usepackage{pifont}
\newcommand{\Full}{\textcolor{green!55!black}{\ding{51}}}
\newcommand{\None}{\textcolor{red!70!black}{\ding{55}}}
\newcommand{\Part}{\CircM}

\usepackage{multirow}
\usepackage{tabularx,booktabs}
\usepackage{longtable,array}

\newcommand{\CircH}{\tikz[baseline=-0.6ex]\fill (0,0) circle (0.55ex);} 
\newcommand{\CircL}{\tikz[baseline=-0.6ex]\draw (0,0) circle (0.55ex);} 
\newcommand{\CircM}{
  \tikz[baseline=-0.6ex]{
    \draw (0,0) circle (0.55ex);
    \path[fill] (0,0) -- (0.55ex,0)
      arc[start angle=0,end angle=180,radius=0.55ex] -- cycle;
  }%
}

\newcommand{\Score}[1]{%
  \ifx#1H\CircH\else%
  \ifx#1M\CircM\else%
  \CircL%
  \fi\fi%
}

\tcbuselibrary{listings,breakable}

\newtcblisting{promptbox}{
  colback=gray!6,
  colframe=gray!40,
  boxrule=0.4pt,
  arc=2pt,
  left=4pt,right=4pt,top=4pt,bottom=4pt,
  listing only,
  listing options={
    basicstyle=\ttfamily\tiny,
    columns=fullflexible,
    breaklines=true,
    breakatwhitespace=true,
    showstringspaces=false,
    literate={≤}{{$\le$}}1 {≥}{{$\ge$}}1 {∈}{{$\in$}}1 {≠}{{$\ne$}}1 {}{{$\bullet$}}1
  },
  breakable,
  enhanced,
}

\newif\ifextended
\extendedfalse

\begin{document}

\date{}

\title{A Graph-Based Analysis of Comparability, Scale, and Attribution
\\in Private Transformer Inference
}

\author{
\IEEEauthorblockN{Souvik Das}
\IEEEauthorblockA{\textit{Texas A\&M University} \\
College Station, TX, USA \\
souvik.das@tamu.edu}
\and
\IEEEauthorblockN{Ariful Azad}
\IEEEauthorblockA{\textit{Texas A\&M University} \\
College Station, TX, USA \\
ariful@tamu.edu}
\and
\IEEEauthorblockN{Nitesh Saxena}
\IEEEauthorblockA{\textit{Texas A\&M University} \\
College Station, TX, USA \\
nsaxena@tamu.edu}
}

\maketitle

\begin{abstract}

Rising privacy concerns around large language models have accelerated research in private Transformer inference (PTI).
Synthesizing the empirical claims in this line of research, however, is hindered by multi-dimensional cryptographic progress and inconsistent evaluation contracts. 
In this work, we systematize the field by constructing an empirical graph of 62 PTI systems, mapping system attributes to node features and evaluation compatibility to edges. 
A three-tiered analysis of this graph illuminates the state of private inference. 
Node-level attribution shows that cryptographic progress drives performance gains across most systems, yet always co-occurs with other factors (model fidelity, hardware, cost accounting). 
Edge-level auditing shows that only 1 of 1891 candidate system pairs is directly comparable under shared assumptions.
Finally, graph-level analysis shows that the scaling frontier of private inference has plateaued at 13B parameters and projects the cost of deploying current PTI techniques on production-scale Transformers.
Overall, this graph-based framework equips the community to evaluate PTI claims transparently and channel technical speedups into deployment-scale private inference.
\end{abstract}


\section{Introduction}
\label{sec:intro}

Large language models (LLMs) increasingly process sensitive user data, ranging from medical records and legal filings to proprietary code~\cite{yan2024protecting}. 
In standard cloud deployment, user inputs and intermediate activations are fully exposed to the model provider. 
Private Transformer Inference (PTI)~\cite{li2024survey,li2025survey} aims to eliminate this exposure by cryptographically protecting client inputs, model weights, or both.

Modern PTI relies on three main cryptographic paradigms: Secure Multi-Party Computation (MPC)~\cite{puma2023,sigma2023,mpcformer2023}, Homomorphic Encryption (HE)~\cite{nexus2024,thex2022,encryptionfriendlyllm2024}, and hybrid MPC--HE constructions~\cite{iron2022,bolt2024,bumblebee2023}.
These paradigms enforce different trade-offs. 
MPC distributes computations interactively across multiple parties, incurring high network round-trips and massive communication volumes. 
Conversely, HE permits non-interactive evaluation over ciphertexts but shifts the burden to local compute and ciphertext expansion. 
To mitigate these costs, recent work emphasizes algorithmic-cryptographic co-designs that leverage polynomial approximations for non-linear layers (e.g., Softmax, GELU)~\cite{mpcformer2023,bolt2024}, SIMD-friendly tensor packing~\cite{nexus2024,iron2022}, dynamic bootstrapping~\cite{cachemir2026}, and protocol-level round reduction~\cite{bumblebee2023,nimbus2024}.
Consequently, reported private inference latencies and communication costs have dropped by orders of magnitude from BERT-scale models~\cite{mpcformer2023,secbert2024} to 13B-parameter decoders~\cite{sigma2023}.

Despite these impressive local gains, collective progress in PTI is remarkably difficult to evaluate. 
A severe disconnect persists between the cryptographic frontier ($\approx$13B parameters) and production models (e.g., LLaMA-3, Kimi K2, GPT-5) operating at hundreds of billions to trillions of parameters~\cite{epochai}. 
More fundamentally, published PTI performance metrics are often incommensurable. 
For example, reported latency may exclude preprocessing, assume high-bandwidth LANs for WAN-motivated use cases, rely on aggressive weight pruning, or protect only input privacy while competing baselines enforce dual privacy. 
Because these unstated variables dictate performance, a result is uninterpretable without its underlying bundle of assumptions, which we term the system's \emph{evaluation contract}. 
Without auditing these contracts side by side, individual improvements cannot be synthesized into a true picture of the state of the field.

\begin{table*}[t]
\centering
\scriptsize
\setlength{\tabcolsep}{3pt}
\renewcommand{\arraystretch}{1.12}
\caption{Comparison between this SoK and prior PTI surveys and systematizations. 
Whereas existing work taxonomizes mechanisms, this work audits whether published evidence is mutually comparable and demonstrates progress toward production-scale private inference.
}
\label{tab:related_work_contrast}
\resizebox{\textwidth}{!}{%
\begin{tabular}{l c c c c c c}
\toprule
\textbf{Work type} &
\textbf{PTI scope} &
\textbf{Mechanism taxonomy} &
\textbf{Repro.\ re-bench.} &
\textbf{Scaling frontier} &
\textbf{Contract comparability} &
\textbf{Progress attribution} \\
\midrule
Private-ML/HE surveys~\cite{falcetta2022he,podschwadt2021sok}
& \Part & \Full & \None & \None & \None & \None \\
Cryptographic NN SoK~\cite{ng2023sok}
& \Part & \Full & \None & \None & \Part & \None \\
Private Transformer surveys~\cite{li2024survey,li2025survey}
& \Full & \Full & \None & \Part & \Part & \Part \\
HE/FHE feasibility SoKs~\cite{gouert2023sok,xue2025sokfullyhomomorphicencryption,pindado2026sok}
& \Part & \Part & \Part & \Part & \None & \None \\
Private Transformer SoK~\cite{chen2026sok}
& \Full & \Full & \Full & \None & \None & \None \\
Approximate-HE SoK, incl.\ PLCS~\cite{al2026sok}
& \Part & \Full & \Part & \Part & \Part & \None \\
\textbf{This work}
& \Full & \Part & \None & \Full & \Full & \Full \\
\bottomrule
\end{tabular}%
}

\vspace{2pt}
{\footnotesize\emph{Legend:} \Full~full support \quad \Part~partial or mechanism-specific \quad \None~not addressed.}
\end{table*}

{\bf Objectives of systematization.}
While PTI research has produced numerous systems, the field lacks a unified empirical framework to determine if reported gains translate to deployment-scale readiness.
An SoK is essential to systematize how evidence is reported and interpreted across conflicting assumptions. 
Existing surveys and SoKs do not address this need since they focus primarily on taxonomizing cryptographic primitives~\cite{ng2023sok, li2024survey, li2025survey}, re-benchmarking select systems in isolated environments~\cite{chen2026sok}, or proposing reporting standards for future work without auditing the record already published~\cite{al2026sok}.  
While valuable, primitive taxonomies cannot explain whether published numbers are methodologically sound to compare, and re-benchmarking misses the broader systemic breakdown in evaluation contracts across the literature.

This SoK directly addresses this gap by auditing the published evidence across 62 PTI papers (2022--2026). 
We reconstruct the evaluation contracts across these papers to determine how many published claims can be directly compared, where comparability breaks down, and what factors truly drive reported speedups. 
Table~\ref{tab:related_work_contrast} summarizes how our evidence-auditing approach complements and extends prior surveys and SoKs.



We center this SoK around four key research questions:

\begin{itemize}[leftmargin=*, topsep=3pt, itemsep=0pt]
  \item \textbf{Comparability (RQ1):} What metadata must a result declare to be interpretable, and how much of the existing literature is directly comparable?
  \item \textbf{Innovation attribution (RQ2):} What factors drive headline performance gains?
  \item \textbf{Scalability and coverage (RQ3):} How far has private inference actually scaled and what cryptographic protection does it cover?
  \item \textbf{Deployment practicality (RQ4):} What is the cost of deployment-scale private inference today, and what would it cost to scale PTI to production-scale (405B) models?
\end{itemize}

{\bf A novel graph-based approach.}
To address these questions, we construct a graph-based framework modeling 62 published PTI systems as nodes and their pairwise evaluation compatibility as edges. 
System specifications, including privacy scope, Transformer architecture, deployment setting, and reporting practices, are encoded as node features. 

To answer RQ1, we evaluate every candidate edge (system pair) through eight compatibility gates. Out of 395 system pairs that share a workload, only a single pair remains directly comparable once fidelity, privacy scope, threat model, cost scope, and reporting standards must also align, and this lone pair represents a deliberate reimplementation on the same benchmark.

To address RQ2, we analyze the graph edges connecting each system to its baselines. These edges reveal the drivers behind performance gains by isolating protocol advances from lighter architectures, nonlinear approximations, batching strategies, hardware differences, and accounting choices. 
While protocol optimizations contribute to reported gains in 49 of 62 papers and confirm genuine cryptographic progress, no headline speedup stems from protocol progress alone. 
Every claim depends on concomitant changes in workload, fidelity, amortization, hardware, or baseline selection.

We answer RQ3 by examining the entire graph to record the largest model evaluated under each paper's privacy assumptions, threat model, and workload. 
Demonstrated coverage concentrates heavily in semi-honest two-party settings on BERT scale encoders. Furthermore, the cryptographic frontier, defined as the largest model evaluated under conventional protection, has plateaued at 13B parameters while production models exceed one trillion, creating a scaling gap of nearly two orders of magnitude.

Finally, to answer RQ4, we fit a cost model within each paper's original contract and project it to larger scale. 
Our projections show that a single private forward pass of a 70B-class model requires minutes to hours of compute time per token, gigabytes of runtime data movement, and terabytes of communication for one-time setup.

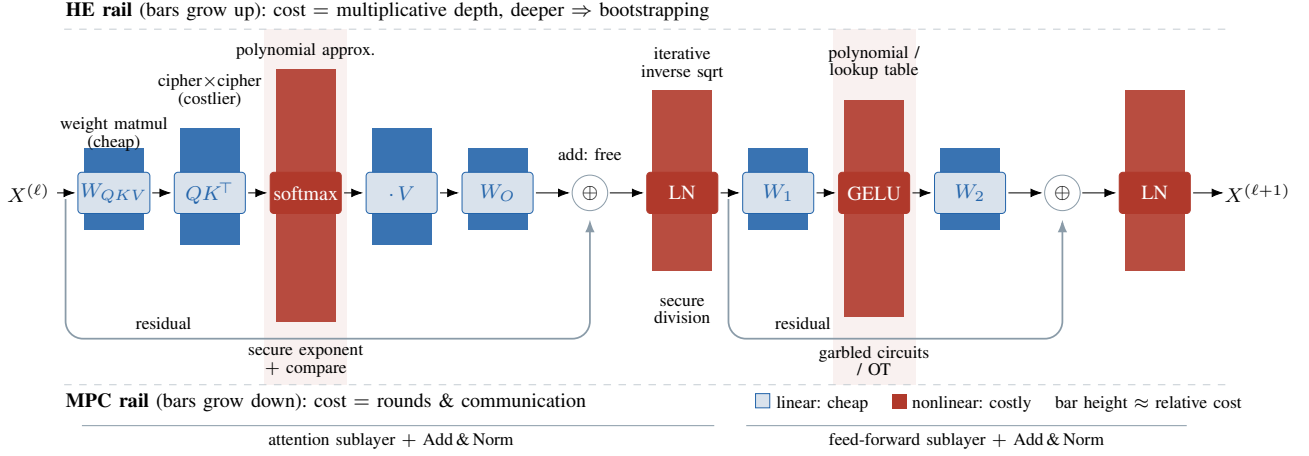
\begin{figure*}[t]
\centering
\definecolor{lincol}{HTML}{2364AA}
\definecolor{linfill}{HTML}{D9E2EC}
\definecolor{noncol}{HTML}{B0392B}
\definecolor{tgray}{HTML}{000000}
\definecolor{resgray}{HTML}{8A9BA8}
\resizebox{\textwidth}{!}{%
\begin{tikzpicture}[font=\footnotesize, yscale=0.72, >=Latex,
  opnon/.style={draw=noncol, fill=noncol, text=white, rounded corners=1.5pt,
                minimum width=1.0cm, minimum height=0.58cm, inner sep=1pt, align=center},
  oplin/.style={draw=lincol, fill=linfill, text=lincol, rounded corners=1.5pt,
                minimum width=1.0cm, minimum height=0.58cm, inner sep=1pt, align=center},
  opadd/.style={draw=resgray, fill=white, text=tgray, circle, inner sep=0pt,
                minimum size=0.5cm},
  io/.style={text=tgray, font=\footnotesize\itshape},
  tip/.style={font=\scriptsize, text=tgray, align=center},
  skip/.style={resgray, line width=0.7pt, rounded corners=7pt, -{Latex[length=1.8mm]}},
]
\begin{scope}[on background layer]
  \fill[noncol!7] (3.42,-3.75) rectangle (4.58,3.2);
  \fill[noncol!7] (11.42,-3.75) rectangle (12.58,3.2);
\end{scope}
\draw[resgray!50, dashed] (0.6,3.2) -- (17.0,3.2);
\draw[resgray!50, dashed] (0.6,-3.75) -- (17.0,-3.75);
\node[tgray, anchor=west] at (0.5,3.55) {\textbf{HE rail} (bars grow up): cost $=$ multiplicative depth, deeper $\Rightarrow$ bootstrapping};
\node[tgray, anchor=west] at (0.5,-4.1) {\textbf{MPC rail} (bars grow down): cost $=$ rounds \& communication};
\node[tgray, anchor=east, font=\scriptsize] at (17.3,-4.1)
  {\tikz\fill[linfill,draw=lincol] (0,0) rectangle (0.2,0.2);~linear: cheap \quad
   \tikz\fill[noncol] (0,0) rectangle (0.2,0.2);~nonlinear: costly \quad
   bar height $\approx$ relative cost};
\newcommand{\anatop}[6]{%
  \fill[#6, fill opacity=0.9] (#1-0.42,0.32) rectangle (#1+0.42,{0.32+#4});
  \fill[#6, fill opacity=0.9] (#1-0.42,-0.32) rectangle (#1+0.42,{-0.32-#5});
  \node[#3] at (#1,0) {#2};
}
\node[io] (in) at (0.1,0) {$X^{(\ell)}$};
\anatop{1.3}{$W_{QKV}$}{oplin}{0.55}{0.4}{lincol}
\anatop{2.65}{$QK^{\!\top}$}{oplin}{0.95}{0.7}{lincol}
\anatop{4.0}{softmax}{opnon}{2.1}{2.2}{noncol}
\anatop{5.35}{$\cdot\,V$}{oplin}{0.95}{0.7}{lincol}
\anatop{6.7}{$W_{O}$}{oplin}{0.55}{0.4}{lincol}
\node[opadd] at (8.0,0) {$\oplus$};
\anatop{9.3}{LN}{opnon}{1.7}{1.2}{noncol}
\anatop{10.65}{$W_{1}$}{oplin}{0.55}{0.4}{lincol}
\anatop{12.0}{GELU}{opnon}{1.5}{2.1}{noncol}
\anatop{13.35}{$W_{2}$}{oplin}{0.55}{0.4}{lincol}
\node[opadd] at (14.65,0) {$\oplus$};
\anatop{15.95}{LN}{opnon}{1.7}{1.2}{noncol}
\node[io] (out) at (17.4,0) {$X^{(\ell+1)}$};
\draw[skip] (0.62,-0.12) |- (4.3,-2.85) -| (8.0,-0.52);
\draw[skip] (9.95,-0.12) |- (12.3,-2.85) -| (14.65,-0.52);
\node[tip] at (2.0,-2.55) {residual};
\node[tip] at (11.0,-2.55) {residual};
\node[tip] at (4.0,2.78) {polynomial approx.};
\node[tip] at (4.0,-3.28) {secure exponent\\[-1pt]$+$ compare};
\node[tip] at (12.0,2.55) {polynomial /\\[-1pt]lookup table};
\node[tip] at (12.0,-3.28) {garbled circuits\\[-1pt]/ OT};
\node[tip] at (9.3,2.52) {iterative\\[-1pt]inverse sqrt};
\node[tip] at (9.3,-2.3) {secure\\[-1pt]division};
\node[tip] at (1.3,1.15) {weight matmul\\[-1pt](cheap)};
\node[tip] at (2.65,2.0) {cipher$\times$cipher\\[-1pt](costlier)};
\node[tip] at (8.0,0.75) {add: free};
\foreach \a/\b in {0.5/0.76, 1.84/2.11, 3.19/3.46, 4.54/4.81, 5.89/6.16,
                   7.24/7.72, 8.28/8.76, 9.84/10.11, 11.19/11.46, 12.54/12.81,
                   13.89/14.37, 14.93/15.41, 16.49/16.92}
  \draw[->, tgray] (\a,0) -- (\b,0);
\node[tgray, font=\scriptsize] at (5.2,-4.85) {attention sublayer $+$ Add\,\&\,Norm};
\node[tgray, font=\scriptsize] at (13.3,-4.85) {feed-forward sublayer $+$ Add\,\&\,Norm};
\draw[resgray] (0.85,-4.55) -- (9.75,-4.55);
\draw[resgray] (10.2,-4.55) -- (16.4,-4.55);
\end{tikzpicture}%
}
\caption{\textbf{Cryptographic cost breakdown of one Transformer encoder layer,}
drawn as in the BERT-family encoders most PTI systems evaluate. Data flows left to right with
the residual bypasses drawn below. The bars above and below each operation sketch
its relative cost on the HE rail (multiplicative depth, up) and the MPC rail
(rounds and communication, down). Weight multiplications (blue) are comparatively
cheap in depth and rounds, less so when the weights are themselves
secret shared, the activation products $QK^{\top}$ and $\cdot V$ cost more,
and the nonlinear steps (red: softmax, GELU, LayerNorm) dominate. The two rails
need not agree: GELU bites hardest in MPC communication but is a shallow
polynomial under HE, while LayerNorm's inverse square root cuts deepest under HE.
This op-level skew is why many reported speedups reshape the nonlinearities
rather than the protocol (RQ2).}
\label{fig:transformer-block}
\end{figure*}

Overall, the paper makes the following contributions:

\begin{enumerate}[leftmargin=*, topsep=3pt, itemsep=0pt]
    \item \textbf{A graph-based framework:} A graph-based framework modeling PTI systems as nodes and their pairwise compatibility as edges. This framework enables a comprehensive analysis of the state of PTI.
    
    \item \textbf{The Private Inference Contract Card:} A standardized reporting scheme defining the essential fields required to make PTI benchmarks interpretable and reproducible. This card is used as node features in the graph.

    \item \textbf{A Comparability Gate Suite:} A set of eight compatibility gates that decide whether two published claims can be compared. These gates are used to label edges in the graph. 

    \item \textbf{A source-of-progress ledger.} We attribute the gain of each system to cryptographic advances and non-cryptographic artifacts (workload, fidelity, hardware, accounting).

    \item \textbf{A scaling frontier for PTI.}
   A year-by-year scaling comparison revealing a two-order-of-magnitude gap between PTI models and frontier LLMs. An empirical cost model then projects deployment overheads to fill this gap.

\end{enumerate}

\section{Background and Preliminaries}
\label{sec:background_preliminaries}

\subsection{Transformer Inference}
A Transformer~\cite{vaswani2017attention}, the basic building block of modern LLMs, stacks tens of identical layers consisting of multi-head attention with feed-forward networks.
Computationally, these layers split into linear matrix multiplications (GEMMs) and fine-grained non-linear operators: Softmax in attention, GELU or ReLU in feed-forward sublayers, and LayerNorm across feature dimensions. 
It is well documented that GEMMs incur the heavy computational burden in plaintext inference~\cite{pope2023efficiently}, but cryptography radically inverts the cost profile~\cite{mpcformer2023}. 
Figure~\ref{fig:transformer-block} illustrates this by mapping the cryptographic overhead on a Transformer layer. 
GEMMs (blue) and residual connections cost little multiplicative depth under HE and a single round under MPC, though GEMM communication still grows with model size.
In contrast, non-linear operators (red) dominate overhead by consuming multiplicative depth or requiring multi-round interaction. 
Driven by this cryptographic cost inversion, the PTI literature overwhelmingly targets non-linearity approximations over core protocol redesigns.

Transformer models come in various flavors: (1)~\emph{Encoder-only} models (e.g., BERT) process fixed-length inputs in a single pass~\cite{mpcformer2023,secbert2024}, (2)~\emph{decoder-only} architectures (e.g., LLaMA) generate tokens autoregressively~\cite{puma2023,sigma2023,nexus2024}, (3)~\emph{encoder--decoder} models condition generation on input sequences~\cite{merge2024,privformer2023}, and (4)~\emph{multimodal} Transformers introduce distinct spatial and textual tensor shapes~\cite{mpcvit2023,salvit2023,privmllm2025}. 
Cryptographic execution costs vary fundamentally across model families, rendering performance metrics directly incomparable across them.

\subsection{Cryptographic Backbones}
PTI frameworks rely on three cryptographic paradigms. 
\emph{Secure Multi-Party Computation (MPC)} distributes secret shares across two or more parties~\cite{puma2023,mpcformer2023,sigma2023}, where linear GEMMs use preprocessed randomness (e.g., Beaver triples) and cost a single online round, while non-linear operators (Softmax, GELU) require multi-round sub-protocols whose communication dominates runtime. 
\emph{Homomorphic Encryption (HE)} schemes (e.g., BFV, BGV, CKKS) evaluate functions non-interactively over ciphertexts~\cite{nexus2024,thex2022,encryptionfriendlyllm2024}, eliminating round complexity at the expense of massive ciphertext expansion, bounded multiplicative depth, and costly bootstrapping. 
\emph{Hybrid HE/MPC} systems integrate both paradigms by executing linear GEMMs under HE and non-linearities under MPC~\cite{iron2022,bolt2024,bumblebee2023}. 
Ultimately, each paradigm pays for privacy along distinct axes: MPC in round trips and network bandwidth, HE in computational density, and hybrids across a combination of both.

\subsection{Protected Assets}


PTI systems protect one of three assets. \emph{Input privacy} hides the client's
input (prompt, document, image) from the model server~\cite{mpcformer2023,puma2023}. \emph{Model privacy} hides the server's
weights and architecture from the client, motivated by providers who treat their
parameters as commercial assets. 
\emph{Mutual privacy} guarantees joint protection for both inputs and model weights. As the most stringent threat model, it incurs the highest cryptographic overhead and serves as the standard paradigm when deploying proprietary models to untrusted clients~\cite{iron2022,bolt2024}.
Even with encrypted inputs and weights, a system can still leak metadata such as sequence length, batch size, token count, or communication volume.
Prior work treats such side-channel leakage inconsistently, ranging from formal security modeling to explicit scope exclusions or omissions.


\subsection{Adversary Models}
A privacy guarantee is meaningful only relative to a threat model. 
Most PTI systems assume a \emph{semi-honest} (passive) adversary that adheres to protocol execution while attempting to infer private state~\cite{bolt2024,bumblebee2023,puma2023}.
By contrast, a \emph{malicious} (active) adversary may arbitrarily deviate from the protocol, incurring substantial cryptographic overhead to mitigate.
Equally critical is the \emph{party model}. Two-party (2PC) protocols compute directly between client and server~\cite{iron2022,bolt2024}, whereas three-party (3PC) or helper-assisted schemes rely on non-colluding entities~\cite{puma2023,sigma2023,shaft2025}. 
Non-colluding helpers reduce non-linear protocol overhead, meaning identical reported execution costs across different party models mask distinct security promises.

\subsection{Deployment Options and Evaluation}
Deployment conditions govern the practical performance of PTI systems, spanning heterogeneous hardware platforms, network topologies, and evaluation regimes. For instance, interactive protocols execute efficiently over high-bandwidth LANs but degrade severely across high-latency WANs, whereas HE comfortably absorbs network round trips~\cite{bolt2024,nexus2024}. 
Similarly, reported overhead varies sharply depending on query granularity (single versus batched query)~\cite{nexus2024}, offline setup reuse (one-time preprocessing versus per-query maintenance)~\cite{sigma2023}, and hardware ranging from single CPUs to multi-GPU clusters~\cite{sigma2023,nexus2024}. 
Finally, empirical reporting across the PTI literature is diverse and often fragmented.  For example, \emph{latency} alternates between online, offline, and end-to-end execution,  \emph{communication} fluctuates across total, per-party, or online-only volume, and interaction \emph{rounds} are selectively disclosed. 
Consequently, the same protocol can substantiate different performance claims depending on its deployment context and evaluation protocols.

\section{Related Work}
\label{sec:related_work}
\paragraph{Surveys and mechanism systematizations.}
Earlier surveys of privacy-preserving machine learning and secure neural-network
computation~\cite{falcetta2022he,podschwadt2021sok,ng2023sok} lay out the
cryptographic building blocks and warn that secure-inference numbers are hard to
compare without watching the assumptions, the network, and the implementation.
Surveys aimed at private Transformer inference~\cite{li2024survey,li2025survey}, a
broader review across protocol, model, and system levels~\cite{zeng2025ppmlreview},
and feasibility SoKs on FHE libraries and
benchmarks~\cite{gouert2023sok,xue2025sokfullyhomomorphicencryption,pindado2026sok}
clarify the design space, but they mostly categorize systems by cryptographic primitives and compiling isolated performance metrics.
Closest to our aims, Al Badawi et al.~\cite{al2026sok} systematize private LLM inference under approximate homomorphic encryption across twenty CKKS frameworks. They observe, as we do, that differences in packing, model fidelity, software stacks, and reported metrics hinder direct comparison, and they propose a Private LLM Card System to standardize how future results are reported; they also place the runtime gap to plaintext at roughly four orders of magnitude. Our PIC card shares that instinct but runs in the opposite direction: instead of specifying what future papers should report, we reconstruct contracts retrospectively from the published record of 62 systems spanning MPC, HE, and hybrid protocols, and use them to measure how much of that record is comparable today. Their systematization sits alongside non-interactive CKKS frameworks such as THOR~\cite{moon2025thor}, MOAI~\cite{zhang2025moai}, and Powerformer~\cite{park2025powerformer}.
We have already compared and contrasted this SoK with prior work in Table~\ref{tab:related_work_contrast}.


{\bf The closest prior systematization.}
The nearest work to ours is a recent SoK on private Transformer inference by Chen et al.~\cite{chen2026sok}, which curates 61 frameworks, categorizes them by cryptographic primitive, and re-benchmarks representative systems under standardized configurations (including WAN environments). Our work and theirs are complementary yet orthogonal. While Chen et al. map and re-measure system \emph{mechanisms} by generating fresh benchmark data, we audit the \emph{published record itself} by evaluating whether reported metrics are mutually comparable and tracking their combined scale. Consequently, rather than introducing another benchmark suite, our contributions establish a field-wide scaling frontier, a standardized Evaluation Contract Card, a pairwise comparability analysis. 
The two studies are complementary, as their benchmarking addresses existing experimental fragmentation, whereas our contract framework keeps future results comparable without requiring continuous re-runs.


{\bf Recent 2026 developments.}
To preserve systematic consistency, our core corpus snapshot was frozen in July 2026, with subsequent literature evaluated as contextual evidence.
We observed three emerging trends in 2026. 
First, private \emph{generation} has surfaced as an explicit target, managing encrypted KV-caches during decoding via hybrid HE--secret sharing~\cite{cryptogen2026} or pure FHE with cache-aware bootstrapping~\cite{cachemir2026}, exposing per-token latency costs that discriminative benchmarks obscure. 
Second, architectural scale is expanding through secure Mixture-of-Experts inference~\cite{secmoe2026}, multi-GPU long-sequence execution~\cite{aegis2026}, hybrid kernel designs~\cite{encformer2026}, and FHE scaled toward 8B parameters~\cite{park2026scaling}. 
Third, non-cryptographic split and permutation schemes have suffered severe attacks, including weight recovery against shuffling defenses~\cite{li2026shuffling} and input reconstruction from split activations~\cite{fan2026splitinference}. 
Other contemporaneous systems, such as truncation-only 2PC~\cite{yubeaton2024truncformer} and speculative-decoding GPTs~\cite{li2025privategpt}, reinforce these same reporting patterns. 
These latest systems sharpen our core thesis that generation length, expert routing, and sequence length introduce new dimensions that no single metric can represent.

\begin{table*}[t]
\centering
\scriptsize
\setlength{\tabcolsep}{3pt}
\renewcommand{\arraystretch}{.9}
\caption{Composition of the audited corpus by cryptographic backbone and evaluated
inference regime. Systems marked [MM] evaluate multiple regimes and appear in both columns. We also audited three encoder-decoder systems: PPTIF~\cite{pptif2024}, Privformer~\cite{privformer2023}, MERGE~\cite{merge2024}. Rhombus~\cite{rhombus2024} is audited at the protocol level: it contributes a matrix-vector multiplication kernel rather than an evaluation in a specific inference regime, so it is not placed in either column.}
\label{tab:taxonomy-ppti}
\resizebox{\textwidth}{!}{%
\begin{tabular}{p{2cm} p{5.8cm} p{5.7cm}}
\toprule
\textbf{Backbone} & \textbf{Encoder} &  \textbf{Decoder} \\
\midrule
\textbf{MPC-based} & MPCFormer~\cite{mpcformer2023}, MPCViT~\cite{mpcvit2023}, RNAViT~\cite{rnavit2023}, SALViT~\cite{salvit2023}, SecBERT~\cite{secbert2024}, SecFormer~\cite{secformer2024}, PrivTF~\cite{privtf2025}, Curl~\cite{curl2024}\,[MM], Ditto~\cite{ditto2024}\,[MM], PUMA~\cite{puma2023}\,[MM], PrivMLLM~\cite{privmllm2025}\,[MM], SHAFT~\cite{shaft2025}\,[MM], SIGMA~\cite{sigma2023}\,[MM], FuseFSS~\cite{fusefss2026}\,[MM], Mosformer~\cite{mosformer2025}\,[MM]  & Comet~\cite{comet2025}, MARILL~\cite{marill2024}, SecureGPT~\cite{securegpt2024}, SecureTLM~\cite{securetlm2024}, Curl~\cite{curl2024}\,[MM], Ditto~\cite{ditto2024}\,[MM], MERGE~\cite{merge2024}\,[MM], PUMA~\cite{puma2023}\,[MM], PrivMLLM~\cite{privmllm2025}\,[MM], SHAFT~\cite{shaft2025}\,[MM], SIGMA~\cite{sigma2023}\,[MM], FLUTE~\cite{flute2026}, MPCache~\cite{mpcache2025}, FuseFSS~\cite{fusefss2026}\,[MM], Mosformer~\cite{mosformer2025}\,[MM] \\
\midrule
\textbf{HE-based} & EncryptionFriendlyLLM~\cite{encryptionfriendlyllm2024}, THEX~\cite{thex2022}, NEXUS~\cite{nexus2024}\,[MM], PowerSoftmax~\cite{powersoftmax2024}\,[MM], AEGIS~\cite{aegis2026}, ATLAS~\cite{atlas2026}\,[MM], ELLMo~\cite{ellmo2026}, Euston~\cite{euston2026}\,[MM], Powerformer~\cite{park2025powerformer}, PriTran~\cite{pritran2026}, THOR~\cite{moon2025thor}  & ENSI~\cite{ensi2025}, EncryptedLLM~\cite{encryptedllm2025}, HEDecoding~\cite{hedecoding2025}, PracticalLLM~\cite{practicalllm2025}, NEXUS~\cite{nexus2024}\,[MM], PowerSoftmax~\cite{powersoftmax2024}\,[MM], ATLAS~\cite{atlas2026}\,[MM], Cachemir~\cite{cachemir2026}, Euston~\cite{euston2026}\,[MM], ScalingPPML~\cite{park2026scaling} \\
\midrule
\textbf{Hybrid MPC--HE} & BOLT~\cite{bolt2024}, CipherFormer~\cite{cipherformer2024}, East~\cite{east2025}, Iron~\cite{iron2022}, LLMEncryptedPrompt~\cite{llmencryptedprompt2023}, Nimbus~\cite{nimbus2024}, Primer~\cite{primer2023}, BLB~\cite{blb2025}\,[MM], BumbleBee~\cite{bumblebee2023}\,[MM], CipherPrune~\cite{cipherprune2025}\,[MM], EncFormer~\cite{encformer2026}\,[MM], SMASH~\cite{smash2026}\,[MM], SecMoE~\cite{secmoe2026}\,[MM]  & AERO~\cite{aero2025}, CipherGPT~\cite{ciphergpt2023}, PermLLM~\cite{permllm2024}, BLB~\cite{blb2025}\,[MM], BumbleBee~\cite{bumblebee2023}\,[MM], CipherPrune~\cite{cipherprune2025}\,[MM], CryptoGen~\cite{cryptogen2026}, CryptoMoE~\cite{cryptomoe2025}, EncFormer~\cite{encformer2026}\,[MM], SMASH~\cite{smash2026}\,[MM], SecMoE~\cite{secmoe2026}\,[MM] \\
\midrule
\textbf{Other / adjacent} & STIP~\cite{stip2023}\,[MM], CASCADE~\cite{cascade2025}\,[MM] & STIP~\cite{stip2023}\,[MM], CASCADE~\cite{cascade2025}\,[MM] \\
\bottomrule
\end{tabular}%
}

\end{table*}

\section{Corpus Construction and Data Collection}
\label{sec:corpus_construction_and_scope}

\paragraph{Scope.}
This SoK covers systems that perform inference on Transformer-based models while keeping the user's input, the model's parameters, or both
hidden behind cryptographic or protocol-level mechanisms. We include a paper when
it (i) targets privacy at inference time, (ii) evaluates Transformer-based or
closely related models such as BERT-style encoders, GPT-style decoders,
encoder--decoder models, ViTs, and multimodal Transformers, (iii) enforces privacy
with MPC, HE, hybrid MPC--HE, secret sharing, or an adjacent protocol-level
mechanism, (iv) presents a complete system or protocol rather than an isolated
primitive, and (v) reports enough evidence to extract at least one model-scale
point, contract field, or progress claim. We exclude training-time methods such as
differential privacy and federated learning, TEE-only systems that rely on hardware
isolation rather than cryptographic protocols, watermarking and ownership schemes,
and cryptographic kernels never integrated into an inference pipeline.

\paragraph{Paper selection.}
We assembled peer-reviewed papers and public preprints published between 2022 and 2026.
We began with keyword searches for ``private Transformer
inference,'' ``secure Transformer inference,'' ``MPC Transformer,'' ``homomorphic
encryption Transformer,'' and ``private LLM inference,'' then followed citations
backward and cross-checked against recent surveys and
systematizations~\cite{li2024survey,li2025survey,ng2023sok,pindado2026sok}. 
We
retain a preprint when it introduces a system-level design that has shaped the
literature or served as a baseline for later work. The final corpus contains 62
papers.
Table~\ref{tab:taxonomy-ppti} shows its composition by cryptographic backbone and inference regime.
This collection was carried out through July 2026 and deliberately folds in the
2025--26 wave of systems (for example CryptoMoE, FLUTE, SMASH, Euston, and THOR),
so that the audit tracks the current frontier rather than a stale snapshot; work
appearing after our collection window is discussed in
Section~\ref{sec:related_work} as context.

\paragraph{Units of analysis.}
Our audit operates on two distinct units of analysis. To map the scaling frontier, the unit is the \emph{paper}, recorded by the largest named model empirically evaluated under a private inference configuration. Models cited solely in motivation or related work are excluded. 
For comparability and innovation attribution, the unit is a \emph{progress claim} paired with its evaluation contract, defined as a reported gain in latency, communication, interaction rounds, scale, or utility against a baseline. 
When a paper reports multiple candidate claims, we select the primary result highlighted in the abstract or main comparison table.

\paragraph{Extraction rules.}
We extract data from each paper based on the following principles. (1) We rely exclusively on published materials, including main text, tables, figures, appendices, and released code artifacts. 
(2) Every extracted attribute is strictly bound to the specific claim under audit, and claims spanning several configurations are designated as multi-setting rather than artificially reduced to a single baseline. 
(3) We resolve ambiguity conservatively by marking missing fields as unknown, classifying ambiguous evaluations as descriptive, and not attributing mixed performance gains to protocol progress. 
(4) We decouple technical merit from comparison validity, recognizing that a technically sound system may still lack an evaluation contract that supports direct comparison.

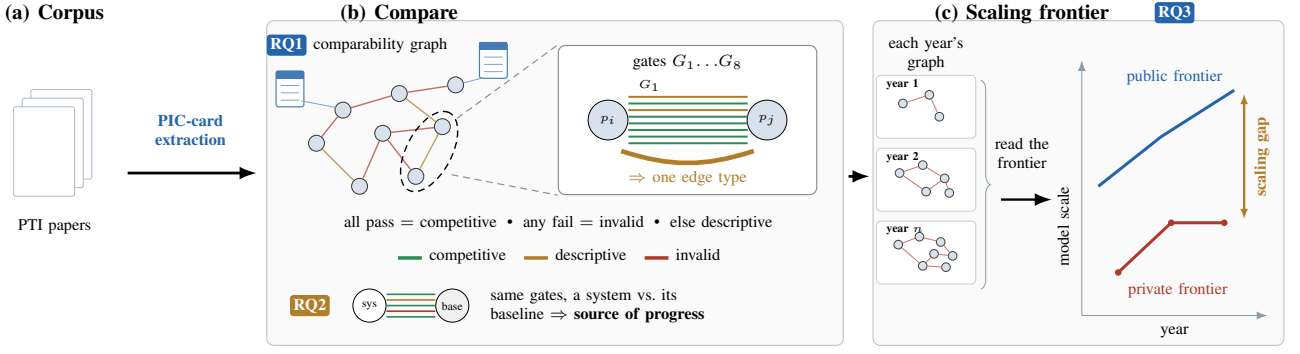
\begin{figure*}[t]
\centering
\definecolor{lincol}{HTML}{2364AA}
\definecolor{linfill}{HTML}{D9E2EC}
\definecolor{noncol}{HTML}{B0392B}
\definecolor{nonfill}{HTML}{F5DEDA}
\definecolor{tgray}{HTML}{000000}
\definecolor{resgray}{HTML}{8A9BA8}
\definecolor{gold}{HTML}{B07D2B}
\definecolor{okgreen}{HTML}{2E8B57}
\resizebox{.99\textwidth}{!}{%
\begin{tikzpicture}[font=\footnotesize, >=Latex,
  panel/.style={draw=resgray!55, rounded corners=3pt, fill=gray!4, line width=0.5pt},
  ptitle/.style={font=\small\bfseries, text=tgray},
  rqtag/.style={font=\scriptsize\bfseries, text=white, fill=lincol, rounded corners=1.5pt, inner sep=1.5pt},
  gnode/.style={draw=tgray!80, fill=linfill, circle, minimum size=0.24cm, inner sep=0pt},
  snode/.style={draw=tgray!75, fill=linfill, circle, minimum size=0.13cm, inner sep=0pt},
  bignode/.style={draw=tgray, fill=linfill, circle, minimum size=0.6cm, inner sep=0pt, font=\tiny, text=tgray},
  flow/.style={-{Latex[length=2.6mm]}, line width=1.2pt, tgray},
  tip/.style={font=\scriptsize, text=tgray, align=center},
  stat/.style={font=\scriptsize, text=tgray, align=center},
]
\def\ptop{5.5}
\def\pcy{2.95}

\newcommand{\minicard}[2]{%
  \fill[white, draw=lincol!85, rounded corners=1pt, line width=0.5pt] (#1-0.21,#2-0.27) rectangle (#1+0.21,#2+0.27);
  \fill[lincol!85] (#1-0.21,#2+0.16) rectangle (#1+0.21,#2+0.27);
  \foreach \dy in {0.05,-0.06,-0.17} {\draw[lincol!60, line width=0.4pt] (#1-0.14,#2+\dy) -- (#1+0.14,#2+\dy);}
}

\def\cx{1.4}
\node[ptitle] at (\cx,\ptop+0.25) {(a) Corpus};
\foreach \i in {0.26,0.13,0} {
  \fill[white, draw=resgray!70, rounded corners=1pt] (\cx-0.62+\i,3.0+\i) rectangle (\cx+0.30+\i,4.35+\i);
}
\node[align=center, text=tgray, font=\scriptsize] at (\cx,2.55) {PTI papers};

\draw[flow] (2.5,3.35) -- (4.45,3.35);
\node[font=\scriptsize\bfseries, text=lincol, align=center] at (3.45,4.0) {PIC-card\\extraction};

\begin{scope}[shift={(-0.6,0)}]
\def\gx{9.55}
\node[panel, minimum width=8.7cm, minimum height=4.9cm] (p3) at (\gx,3.2) {};
\node[ptitle] at (\gx-2.35,\ptop+0.25) {(b) Compare};

\node[rqtag] at (5.5,5.28) {RQ1};
\node[tip, anchor=west] at (5.78,5.28) {comparability graph};
\foreach \n/\x/\y in {a/6.35/4.30, b/7.20/4.55, c/7.85/4.05, d/7.45/3.30, e/6.60/3.15, f/5.95/3.80, g/6.95/3.95, h/8.05/4.68} {
  \node[gnode] (\n) at (\x,\y) {};
}
\foreach \u/\v/\c in {a/b/noncol, b/c/gold, c/d/gold, e/g/noncol, g/d/noncol, b/h/noncol, f/a/noncol, f/e/gold, g/c/noncol} {
  \draw[\c!75, line width=0.6pt] (\u) -- (\v);
}
\minicard{5.55}{4.62}
\draw[lincol!60, line width=0.4pt] (5.72,4.44) -- (6.24,4.34);
\minicard{8.62}{5.05}
\draw[lincol!60, line width=0.4pt] (8.47,4.79) -- (8.17,4.71);

\coordinate (lensc) at (7.65,3.675);
\draw[tgray, densely dashed, line width=0.6pt, rotate around={62:(7.65,3.675)}] (lensc) ellipse (0.66 and 0.34);

\node[panel, fill=white, draw=tgray!45, minimum width=3.9cm, minimum height=2.25cm] (zoom) at (11.55,4.15) {};
\draw[tgray!45, densely dashed, line width=0.5pt] ($(lensc)+(0.42,0.5)$) -- (zoom.north west);
\draw[tgray!45, densely dashed, line width=0.5pt] ($(lensc)+(0.34,-0.35)$) -- (zoom.south west);
\node[bignode] (A) at (10.35,4.15) {$p_i$};
\node[bignode] (B) at (12.75,4.15) {$p_j$};
\foreach \k/\col in {1/okgreen,2/okgreen,3/okgreen,4/okgreen,5/okgreen,6/gold,7/okgreen,8/gold} {
  \draw[\col, line width=0.85pt] ($(A.east)+(0,{(\k-4.5)*0.1})$) -- ($(B.east)+(-0.6,{(\k-4.5)*0.1})$);
}
\node[font=\tiny,text=tgray] at ($(A.east)+(0.32,0.52)$) {$G_1$};
\node[tip] at (11.55,5.02) {gates $G_1\!\dots\!G_8$};
\draw[gold, line width=2.0pt] ($(A.south)+(0.2,-0.16)$) to[out=-18,in=198] ($(B.south)+(-0.2,-0.16)$);
\node[tip,text=gold] at (11.55,3.28) {$\Rightarrow$ one edge type};

\node[tip] at (\gx+0.05,2.6) {all pass $=$ competitive \;\textbullet\; any fail $=$ invalid \;\textbullet\; else descriptive};
\node[stat] at (\gx+0.05,2.1) {%
  \tikz\draw[okgreen,line width=1.5pt](0,0)--(0.36,0); competitive\quad
  \tikz\draw[gold,line width=1.5pt](0,0)--(0.36,0); descriptive\quad
  \tikz\draw[noncol,line width=1.5pt](0,0)--(0.36,0); invalid};

\node[rqtag, fill=gold] at (5.85,1.35) {RQ2};
\node[bignode, fill=white, minimum size=0.5cm] (S) at (6.75,1.35) {\tiny sys};
\node[bignode, fill=gray!12, minimum size=0.5cm] (BL) at (8.0,1.35) {\tiny base};
\foreach \k/\col in {1/okgreen,2/noncol,3/okgreen,4/gold,5/okgreen} {
  \draw[\col, line width=0.7pt] ($(S.east)+(0,{(\k-3)*0.085})$) -- ($(BL.west)+(0,{(\k-3)*0.085})$);
}
\node[stat, anchor=west, align=left] at (8.45,1.35) {same gates, a system vs.\ its\\baseline $\Rightarrow$ \textbf{source of progress}};
\end{scope}

\def\ox{16.85}
\node[panel, minimum width=6.2cm, minimum height=4.9cm] (p4) at (\ox,3.2) {};
\node[ptitle] at (\ox-0.85,\ptop+0.25) {(c) Scaling frontier};
\node[rqtag] at (\ox+1.45,\ptop+0.27) {RQ3};

\node[tip] at (14.55,5.15) {each year's\\graph};
\newcommand{\yeargraph}[3]{%
  \node[draw=resgray!50, fill=white, rounded corners=2pt, minimum width=1.5cm, minimum height=0.95cm] at (#1,#2) {};
  \node[font=\tiny\bfseries, text=tgray, anchor=west] at (#1-0.72,#2+0.34) {#3};
}
\yeargraph{14.55}{4.35}{year 1}
\foreach \n/\x/\y in {A/-0.35/0.05, B/0.05/0.18, C/0.2/-0.15} {\node[snode](t22\n) at (14.55+\x,4.35+\y){};}
\draw[noncol!70,line width=0.4pt] (t22A)--(t22B) (t22B)--(t22C);
\yeargraph{14.55}{3.25}{year 2}
\foreach \n/\x/\y in {A/-0.4/0.1, B/-0.05/0.22, C/0.28/0.02, D/0.0/-0.18, E/0.35/-0.2} {\node[snode](t24\n) at (14.55+\x,3.25+\y){};}
\draw[noncol!70,line width=0.4pt] (t24A)--(t24B) (t24B)--(t24C) (t24D)--(t24C) (t24A)--(t24D) (t24C)--(t24E);
\yeargraph{14.55}{2.15}{year $n$}
\foreach \n/\x/\y in {A/-0.42/0.12, B/-0.1/0.24, C/0.22/0.16, D/0.4/-0.05, E/0.12/-0.02, F/-0.05/-0.2, G/0.3/-0.22} {\node[snode](t26\n) at (14.55+\x,2.15+\y){};}
\draw[noncol!70,line width=0.4pt] (t26A)--(t26B) (t26B)--(t26C) (t26C)--(t26D) (t26E)--(t26F) (t26E)--(t26D) (t26F)--(t26G) (t26A)--(t26F);

\draw[decorate, decoration={brace, amplitude=4pt}, resgray] (15.38,4.75) -- (15.38,1.6);
\draw[flow, line width=1.0pt] (15.68,2.95) -- (16.45,2.95);
\node[tip, text=tgray] at (16.0,3.65) {read the\\frontier};

\def\px{16.9}
\draw[resgray,-{Latex[length=1.6mm]}] (\px,1.2) -- (\px,5.05);
\draw[resgray,-{Latex[length=1.6mm]}] (\px,1.2) -- (19.7,1.2);
\node[font=\scriptsize,text=tgray,rotate=90] at (\px-0.28,2.55) {model scale};
\node[font=\scriptsize,text=tgray] at (18.3,0.95) {year};
\draw[lincol, line width=1.3pt] (17.15,3.15) -- (18.1,3.9) -- (19.2,4.6);
\node[font=\scriptsize,text=lincol] at (18.3,4.8) {public frontier};
\draw[noncol, line width=1.3pt] (17.45,1.85) -- (18.25,2.6) -- (19.05,2.6);
\fill[noncol] (17.45,1.85) circle (1.4pt);
\fill[noncol] (18.25,2.6) circle (1.4pt);
\fill[noncol] (19.05,2.6) circle (1.4pt);
\node[font=\scriptsize,text=noncol] at (18.35,1.58) {private frontier};
\draw[{Latex[length=1.6mm]}-{Latex[length=1.6mm]}, gold, line width=1.0pt] (19.35,2.65) -- (19.35,4.55);
\node[font=\scriptsize\bfseries,text=gold,rotate=90] at (19.62,3.6) {scaling gap};

\draw[flow] (13.38,3.3) -- (13.7,3.3);

\end{tikzpicture}%
}
\caption{\textbf{An overview of the graph-based framework.} \textbf{(a)}~Every paper in the
corpus enters the audit as a Private Inference Contract (PIC) card.
\textbf{(b)}~Each pair of cards is evaluated through the comparability gates
$G_1$--$G_8$, which aggregate into a single edge type in the comparability graph
(RQ1); the same gates, read against a paper's own baseline, identify its source of
progress (RQ2). \textbf{(c)}~Reading the per-year graphs together traces the
private scaling frontier against the public one (RQ3).}
\label{fig:method-overview}
\end{figure*}

\paragraph{Data validation.}
\label{sec:method_validation}
Because the values of record are extracted by hand, we stress-test them with an
independent blind re-extraction: for each paper a separate large-language-model
reader (GPT-5.6-Luna) re-derives every contract-card field from the source text
alone, blind to the recorded value, and we compare the two readings field by field.
The model is only a second independent reader of published text and never produces a
value of record, so what we measure is whether the recorded values reproduce from
the source, not whether a model could replace the human auditor. Agreement is
strong on exactly the fields the gates lean on most (87--98\% on cryptographic
backbone, privacy scope, adversary model, and end-to-end status), with 70.0\%
raw agreement pooled across all factual fields. We adjudicated every flagged
disagreement against the source papers: most trace to conversion losses or to
values the blind reader declines to infer, and, across the two adjudication
passes, 40 of 992 fields (4.0\%) were corrected in the corpus.
The full per-field breakdown is in Appendix~\ref{app:verification}
(Table~\ref{tab:verification-agreement}).

\section{Systematization Methodology}
\label{sec:evaluation_methodology}
This section establishes an analytical framework to synthesize the cryptographic and evaluation contexts across all 62 papers in our corpus. 
As illustrated in Figure~\ref{fig:method-overview}, each paper is first reduced to a structured record, termed a \emph{PIC card}, that captures its performance claims and underlying operating assumptions. From these cards, we construct a pairwise comparability graph where system pairs are connected across a suite of evaluation \emph{gates}. Together, these cards and gates enable analysis at three distinct levels: (1)~evaluating cross-system comparability to answer RQ1; (2)~isolating the precise drivers of reported speedups to answer RQ2; and (3)~tracing the field's true scaling trajectory longitudinally over time. The remainder of this section details each stage of this pipeline.

\subsection{The Private Inference Contract (PIC) Card}
\label{sec:method_contract_card}
If we treat the corpus of PTI papers as a database of reported results, its entities cannot be meaningfully compared or aggregated without a schema defining what each field means and under what conditions it holds.  
For example, two papers reporting ``latency'' may differ in threat models, preprocessing costs, or approximations, and none of this is visible from the number itself.
We formalize the Private Inference Contract (PIC) card as a unified schema defining the essential fields required to interpret an individual PTI result and evaluate it side by side with another.

{\bf Structure of the PIC card.} 
To reflect the multidimensional nature of PTI systems, the card is organized into five \emph{categories}, each containing several related \emph{fields}. 
Table~\ref{tab:evaluation-contract-card} details these five categories and their constituent fields.
The \emph{scale and workload} group defines the exact computation performed, recognizing that an autoregressive 7B decoder generating individual tokens evaluates a fundamentally different workload than a 110M encoder classifying fixed-length sequences.
The \emph{privacy and threat model} group establishes the security guarantees provided, distinguishing a two-party semi-honest protocol from a three-party topology with a non-colluding helper that may yield identical latencies under fundamentally different trust assumptions. Model privacy here means the weights: architecture, tensor dimensions, and sequence length usually remain visible. 
The \emph{mechanism and fidelity} group separates cryptographic advances from structural modifications to the underlying model, ensuring that speedups from polynomial approximations or weight pruning are not misattributed to protocol optimizations. 
The \emph{cost and deployment} group specifies the operational measurement conditions, differentiating end-to-end latencies from online only execution, LAN from WAN network topologies, and single-query measurements from amortized batch figures. 
Finally, the \emph{baseline and reporting} group documents comparison baselines and field completeness, establishing whether published evidence is sufficient to enable a rigorous comparison.


\begin{table*}[t]
\centering
\small
\setlength{\tabcolsep}{5pt}
\renewcommand{\arraystretch}{1.12}
\caption{Evaluation contract card used to interpret PTI progress claims. A
headline number is directly comparable only when the relevant contract fields are
aligned and sufficiently reported.}
\label{tab:evaluation-contract-card}
\begin{tabularx}{\textwidth}{p{3.3cm} X}
\toprule
\textbf{Contract category} & \textbf{Fields audited} \\
\midrule
Scale and workload &
Largest evaluated model, parameter count, inference mode (encoder / decoder /
encoder--decoder) \\

Privacy and threat model &
Protected asset, leakage scope, adversary model, party model \\

Mechanism and fidelity &
Mechanism (MPC / HE / hybrid), fidelity class (exact, approximate,
architecture-modified, task-limited) \\

Cost and deployment &
Latency, communication, rounds, cost scope (online / offline / end-to-end),
amortization, network regime \\

Baseline and reporting &
Named baseline, in-corpus or external baseline, missing or limiting fields,
sufficiency for cross-paper comparison \\
\bottomrule
\end{tabularx}
\end{table*}

{\bf Construction of a PIC card for a paper.}
We fill one card per paper by manual extraction under the rules of Section~\ref{sec:corpus_construction_and_scope}. A human auditor records a field only when the manuscript states it explicitly, following the codebook in Appendix~\ref{app:llm_protocol}, and an independent LLM re-extraction verifies the recorded values (Section~\ref{sec:method_validation}).
When a field is absent or ambiguously stated in a paper, we record it as \emph{unknown}. This classification is not a penalty, but a preserved signal that the published evidence cannot support direct comparison along that dimension.

{\bf Example of a PIC card.} Figure~\ref{fig:pic-instantiation} demonstrates the card instantiated for BumbleBee~\cite{bumblebee2023}, a state-of-the-art PTI system.
The card explicitly captures the privacy contract, Transformer architecture, and experimental context used to evaluate BumbleBee. 
Hence, the card provides the necessary operational context to interpret performance, contextualizing performance numbers in ways an isolated execution time cannot.
Side-by-side card comparisons expose where system assumptions align or diverge across the literature, forming the foundation for the comparability gates formalized in the next subsection.




\begin{figure}[t]
\centering
\definecolor{lincol}{HTML}{2364AA}
\definecolor{linfill}{HTML}{D9E2EC}
\newcommand{\chip}[1]{\tcbox[on line, nobeforeafter, colback=linfill, colframe=linfill,
  boxsep=1pt, left=3pt, right=3pt, top=1pt, bottom=1pt, arc=2pt]{\textcolor{lincol}{\bfseries #1}}}
\newcommand{\dsep}{\par\vspace{3pt}{\color{lincol!30}\hrule height 0.3pt}\vspace{3pt}}
\begin{tcolorbox}[width=\columnwidth, colback=white, colframe=lincol, boxrule=1pt, arc=3pt,
  left=5pt, right=5pt, top=3pt, bottom=4pt,
  title=\textbf{PIC Card: BumbleBee}~\textnormal{(example instantiation)},
  fonttitle=\footnotesize, coltitle=white, colbacktitle=lincol]
\footnotesize\setlength{\parindent}{0pt}
\chip{Scale \& workload}\; LLaMA-7B (7B) largest; also BERT, GPT-2, ViT; enc.\ + dec.\ [MM]
\dsep
\chip{Privacy \& threat}\; input + model; semi-honest; 2PC
\dsep
\chip{Mechanism \& fidelity}\; hybrid MPC--HE; approximate model
\dsep
\chip{Cost \& deployment}\; end-to-end; LAN + WAN; not amortized
\dsep
\chip{Baseline \& reporting}\; latency \Full, comm.\ \Full, rounds \Part\ (component-level)
\end{tcolorbox}
\caption{The PIC card instantiated for BumbleBee~\cite{bumblebee2023}, filled from
the source paper through the card in Table~\ref{tab:evaluation-contract-card}. A check marks a fully reported field and a half-circle a partially reported one
(here, rounds only at the component level).}
\label{fig:pic-instantiation}
\end{figure}

\begin{table*}[t]
\centering
\small
\caption{Refinement of the five PIC card categories into eight comparability gate types.}
\label{tab:gate-to-section}
\begin{tabularx}{\textwidth}{p{2.2cm} p{3.2cm} X}
\toprule
\textbf{Gate} & \textbf{PIC category} & \textbf{Evaluation predicate} \\
\midrule
G1: Privacy & Privacy and threat model & Protected assets and leakage scope match \\
G2: Threat & Privacy and threat model & Adversary model and party topology match \\
G3: Workload & Scale and workload & Evaluated model architecture and workload match (e.g., encoder vs decoders)\\
G4: Fidelity  & Mechanism and fidelity & Model preservation or approximation degree matches \\
G5: Cost & Cost and deployment & Cost accounting boundaries and execution phases match (e.g., end-to-end vs online-only) \\
G6: Amortization & Cost and deployment & Batch and sequence amortization assumptions match \\
G7: Network & Cost and deployment & Network bandwidth and latency environments match (e.g., LAN vs WAN)\\
G8: Reporting & Baseline and reporting & Report sufficient context for comparison (e.g., latency, communication, and rounds) \\
\bottomrule
\end{tabularx}
\end{table*}


\subsection{Comparability Graph and Gates }
\label{sec:method_gates}
\label{sec:comparability_graph}
We model the audited corpus as a heterogeneous multigraph $\mathcal{G} = (V, E)$, where each vertex $v \in V$ represents a paper defined by its PIC card. 
Because published systems evaluate disparate architectures, threat models, and accounting regimes, no single predicate can determine overall comparability. Instead, two papers $a$ and $b$ are connected by multiple typed edges, each representing a specific dimension of evaluation compatibility. A typed edge exists if and only if both papers satisfy compatibility criteria along that dimension.

{\bf Refining PIC categories into comparability gates.}
To provide diagnostic resolution, we refine the five PIC card categories into eight comparability gates $G_1, \ldots, G_8$, corresponding to distinct edge types in $\mathcal{G}$. The map is one to many: a category with several independently comparable fields contributes several gates (privacy and threat into $G_1$--$G_2$, cost and deployment into $G_5$--$G_7$), while the rest contribute one each. A gate $G_i(a,b)$ evaluates a predicate over the corresponding fields, returning $\mathrm{pass}$ when they match, $\mathrm{fail}$ when values explicitly conflict, and $\mathrm{unknown}$ when required metadata is omitted from either paper. Table~\ref{tab:gate-to-section} lists the eight gates and their evaluation predicates.

{\bf Aggregating into a single comparability label.}
We determine aggregate pair comparability by evaluating the conjunction $L(a,b) = \bigwedge_{i=1}^{8} G_i(a,b)$ under three valued logic ($\mathrm{fail} < \mathrm{unknown} < \mathrm{pass}$). 
A single failing gate forces the aggregate label to $\mathrm{fail}$, while missing fields yield $\mathrm{unknown}$ in the absence of explicit failure. Collapsing the multigraph via $L(a,b)$ partitions paper pairs into three operational relations: \emph{competitive} ($L = \mathrm{pass}$, indicating complete contract alignment for direct headline comparison), \emph{descriptive} ($L = \mathrm{unknown}$, where incomplete reporting prevents verification), and \emph{invalid} ($L = \mathrm{fail}$, where explicit contract incompatibility renders direct comparison misleading). 
These labels evaluate comparison validity strictly and do not reflect technical merit.

{\bf A worked example.}
To illustrate the operational mechanics of the comparability gates, consider BOLT~\cite{bolt2024} and BumbleBee~\cite{bumblebee2023}, two systems frequently compared in the literature  when evaluating BERT-base under two-party hybrid HE and MPC.
Although superficially comparable, walking the gate hierarchy reveals subtle contract divergences. Six gates pass: privacy scope ($G_1$), threat model ($G_2$), workload ($G_3$, since both evaluate BERT-base), model fidelity ($G_4$, both approximate), cost scope ($G_5$), and network regime ($G_7$).
The pair still does not reach a competitive relation, but for lack of reporting rather than outright conflict: amortization ($G_6$) is unknown because BOLT does not state whether its cost is amortized, and reporting sufficiency ($G_8$) is unknown because BumbleBee reports rounds only at the component level. With no explicit contract incompatibility but two fields left unresolved, the gate conjunction places these systems in a descriptive relationship rather than a direct competitive ranking, a comparison that can be described but not certified.

\subsection{Progress Attribution}
\label{sec:method_attribution}

A reported performance gain reflects a relational comparison between a progress claim $c$ and a baseline configuration $c_0$.
Because both $c$ and $c_0$ are fully instantiated as PIC cards, progress attribution reduces to identifying contract field discrepancies across the pair. 
We formalize this process through the operational outcomes of the eight comparability gates evaluated on $(c, c_0)$, isolating genuine cryptographic progress from compound gains driven by non-cryptographic factors.

\begin{enumerate}[leftmargin=*, topsep=3pt, itemsep=0pt]
    \item \textbf{Protocol progress:} No non-cryptographic gate is in explicit conflict, and the \emph{dominant} discrepancy between $c$ and $c_0$ is the cryptographic primitive or subprotocol mechanism; any residual non-protocol difference is recorded as a secondary caveat.
    
    \item \textbf{Assumption tradeoff:} Triggered when privacy scope ($G_1$) or threat model ($G_2$) fails, indicating a relaxed adversary model, altered party topology, or expanded trust setup.
    
    \item \textbf{Workload or fidelity shift:} Triggered when scale and workload ($G_3$) or mechanism and fidelity ($G_4$) fails, indicating altered model architectures, sequence lengths, pruning, quantization, or non linear approximations.
    
    
    \item \textbf{Hardware or deployment shift:} Triggered when network regime ($G_7$) fails or underlying hardware specifications differ between $c$ and $c_0$.
    
    \item \textbf{Component progress:} Represents a structural scope mismatch where workload ($G_3$) and cost scope ($G_5$) evaluate an isolated subprotocol or operator rather than an end to end inference system.
    
    \item \textbf{Reporting artifact:} Triggered when reporting sufficiency ($G_8 {=}\mathrm{unknown}$) prevents verification of baseline metadata.
\end{enumerate}
When evaluating $(c, c_0)$ yields $\mathrm{fail}$ or $\mathrm{unknown}$ outcomes across multiple gates, the claim receives a \textbf{mixed attribution} label, recording all contributing gate discrepancies as secondary caveats. 
This formulation enforces a strict invariant: a claim receives a pure protocol progress label if and only if every non cryptographic gate evaluates to $\mathrm{pass}$. 
Any co-occurring gate discrepancy disqualifies the claim from \emph{pure} protocol progress: it is recorded as a secondary caveat and, when no single driver dominates, the claim is labeled mixed attribution. 
As detailed in Section~\ref{sec:results_progress_attribution}, while 49 of 62 audited papers demonstrate genuine cryptographic progress, not a single headline speedup stems from protocol advances alone.

\subsection{Scaling Frontier Audit}
\label{sec:method_scaling}

The preceding sections establish three lenses on the corpus where the PIC card describes what a single paper claims, the gates describe whether two claims can be placed side by side, and attribution describes what kind of change produced a given gain. 
Taken together, over all 62 nodes of $\mathcal{G}$, they support a claim none of the three supports alone, namely where the field's demonstrated capability actually stands.

\label{sec:method_scaling}
To evaluate the scaling gap separating PTI from unprotected LLMs, we benchmark the private inference parameter envelope against unencrypted production models. 
Let $P(v)$ denote the maximum parameter count reported in the PIC card for paper $v$. 
We define the PTI scaling frontier $P_{\mathrm{PTI}}(y)$ as the largest protected model published by year $y$:
\begin{equation}
P_{\mathrm{PTI}}(y) = \max \Big\{ P(v) \;\Big|\; v \in V_{\mathrm{conv}}, \; \text{Year}(v) \le y \Big\}
\end{equation}
where $V_{\mathrm{conv}} \subseteq V$ represents the subset of articles that evaluate conventional cryptographic implementations, filtered through gates $G_4$ (model fidelity) and $G_5$ (cost scope).
We benchmark this against $P_{\mathrm{LLM}}(y)$, the parameter envelope of unencrypted production transformer models published by year $y$. 

\section{Audit Findings}
\label{sec:empirical_results}

This section presents our audit findings addressing RQ1 through RQ3, leaving RQ4 for Section~\ref{sec:cost_roadmap}. 
By synthesizing comparability graphs alongside each node's PIC card, we demonstrate the SoK's central thesis: direct cross-system comparability remains severely constrained, and  there is a widening scaling gap between PTI frameworks and state-of-the-art plaintext models.

\subsection{Overview of PIC Cards}
\label{sec:results_graph_anatomy}
Before addressing our research questions, we synthesize the overall landscape of the corpus using its extracted PIC cards and comparability gates. 
Figure~\ref{fig:pic-distributions} in  Appendix~\ref{app:pic_distributions} details how the 62 audited papers distribute across key system fields.
While the corpus exhibits strong consensus around threat models (55 semi-honest, 4 malicious, 3 unclear) and privacy scopes (49 protecting both input and model), it spans broad structural diversity across remaining axes: cryptographic mechanism (24 MPC, 17 HE, 19 hybrid), model fidelity (40 approximate, 14 modified, 7 exact), party topology (27 2PC, 16 client-server, 6 multi-server, 6 helper, 7 other), workload regime (21 encoder-only, 21 both, 17 decoder-only), and reported cost scope (42 end-to-end, 11 online-only, 9 offline/narrower).
This structural divergence makes the comparability graph sparse, constraining direct cross-paper evaluations as detailed in the following sections.

\begin{figure}[t]
\centering
\includegraphics[width=\columnwidth]{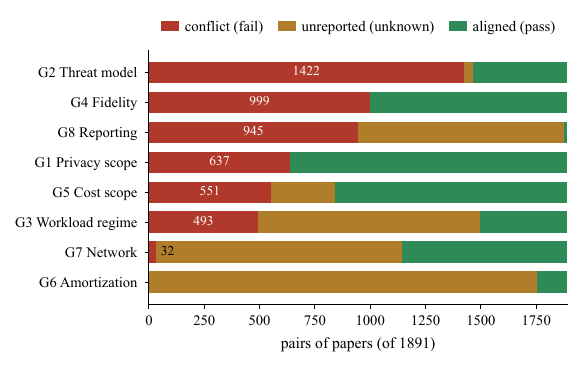}
\caption{Per-gate outcome across all system pairs. The threat and party model
(G2) and model fidelity (G4) drive most of the hard incompatibility, while
amortization (G6) and reporting sufficiency (G8) are dominated by unreported fields.}
\label{fig:gate-breakdown}
\end{figure}

\begin{figure}[t]
\centering
\includegraphics[width=\columnwidth]{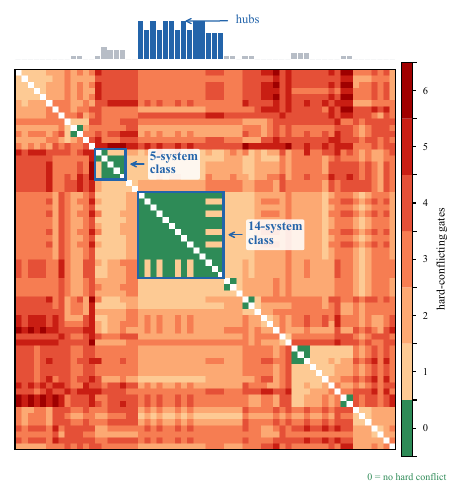}
\caption{
Pairwise comparability matrix across the 62 audited papers. Cell color indicates the number of hard-conflicting gates (green: no conflict; darker red: higher conflict). 
Rows and columns are ordered via agglomerative clustering, with boxed blocks marking the two comparable cohorts and the top bar tracking each paper's count of conflict-free partners. See Figure~\ref{fig:comparability-heatmap-full} in Appendix~\ref{app:comparability_matrix} for the fully labeled matrix.}
\label{fig:comparability-heatmap}
\end{figure}

\begin{figure}[t]
\centering
\includegraphics[width=\columnwidth]{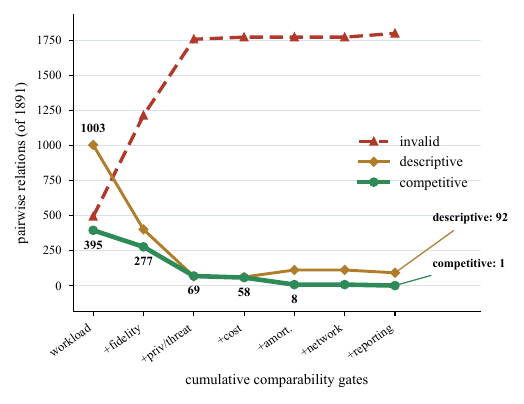}
\caption{The comparability collapse under cumulative gates (RQ1). Competitive
pairs fall from 395 under a compatible workload alone to a single pair under the
full contract, with most of the drop at the early substantive gates rather than at
reporting; invalid pairs rise to match, and descriptive pairs are the residual.}
\label{fig:gate-ablation}
\end{figure}

\subsection{RQ1: Comparability Between Systems}
\label{sec:results_comparability}

{\bf Pairwise comparability.}
Figure~\ref{fig:gate-breakdown} summarizes pairwise agreement and conflict across all eight comparability gates.
Hard conflicts dominate threat and party models (G2: 1422 of 1891 pairs) and model fidelity (G4: 999 pairs), meaning a random pair typically disagrees on trust assumptions or model integrity. 
Conversely, amortization (G6) and reporting sufficiency (G8) are bottlenecked by unreported parameters rather than explicit conflicts. 

Figure~\ref{fig:comparability-heatmap} illustrates pairwise conflict density across all 1891 pairs, with comparable systems clustered along the diagonal. We observe only two distinct, relatively comparable cohorts: (1) a 14-system cluster (BLB, BOLT, BumbleBee, CipherGPT, CryptoMoE, East, EncFormer, Euston, FLUTE, MPCFormer, NEXUS, Nimbus, SHAFT, SecMoE) and (2) a 5-system cluster (AERO, CipherFormer, CipherPrune, LLMs-Can-Understand-Encrypted-Prompt, MPCViT). The remaining corpus is heavily fragmented, with 32 papers sharing zero conflict-free pairings. As highlighted by the top bar of Figure~\ref{fig:comparability-heatmap}, the most connected systems (BLB, BumbleBee, EncFormer, Euston, NEXUS, SHAFT, and SecMoE) share no hard conflict with about~20\% of published works, explaining why they appear most frequently as baselines in prior literature.

{\bf Cumulative comparability.}
Comparability across PTI systems degrades rapidly under cumulative gate enforcement (Figure~\ref{fig:gate-ablation}). 
Isolated regime matching yields 395 compatible pairs (20.9\%), but layering successive contract constraints rapidly erodes this pool. 
Model fidelity drops valid pairs to 277 (14.6\%), threat model, privacy, and cost narrow the count to 58 (3.1\%). Factoring in other gates isolates a single fully compliant pair. In total, cumulative gate filtering invalidates 1798 of 1891 candidate pairings (95.1\%), leaving just 92 descriptive and 1 competitive pair.
The sole pair surviving all gates succeeds only because one system explicitly reimplemented the other.

Two checks prove this collapse reflects true contract incompatibility rather than missing data. 
First, hard fails eliminate 339 of 395 workload-compatible pairs (fidelity removing 118 and threat models removing 208) compared to just 55 unknown failures. Second, gate order and single-gate relaxations do not change the outcome because threat models (1422/1891) and fidelity (999/1891) independently fail most pairs (Figure~\ref{fig:gate-breakdown}).
We emphasize that we can still benchmark systems under shared contracts, but current published evidence cannot support a global leaderboard.

\begin{tcolorbox}[
  enhanced,
  colback=blue!5,
  colframe=black,
  boxrule=0.5pt,
  arc=5pt,
  left=2pt, right=2pt, top=2pt, bottom=2pt,
  colbacktitle=gray!10,
  coltitle=black,
  fonttitle=\bfseries
]
{\bf Finding 1:} Comparability in PTI is not just rare (1 competitive pair out of 1891) but structurally concentrated in a small hub of systems, leaving most of the corpus unable to be compared to anything at all.
\end{tcolorbox}

\begin{figure}[t]
\centering
\includegraphics[width=\columnwidth]{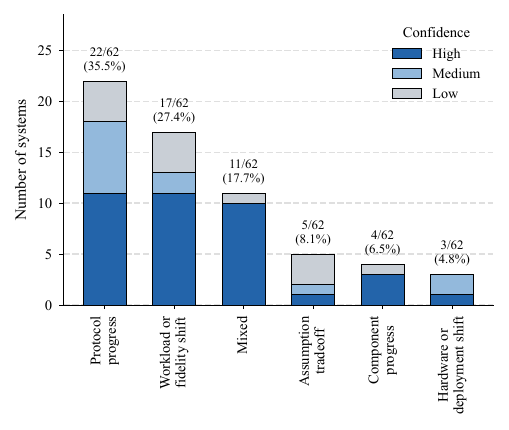}
\caption{Primary progress-attribution labels over the 62-paper corpus. These labels attribute the main source of each paper's reported progress claim; a primary protocol label marks the dominant driver, and every such paper still carries at least one secondary non-protocol caveat (Section~\ref{sec:results_progress_attribution}).
Each bar is split by our confidence in the assigned label: \emph{high} when the paper names its baseline and states the changed component explicitly, \emph{medium} when the label is clear but carries a stated secondary caveat, and \emph{low} when the label requires inference from indirect evidence or sits near a category boundary. Dropping the low-confidence labels leaves the same findings (Appendix~\ref{sec:results_sensitivity}).
}
\label{tab:progress-attribution}
\end{figure}


\subsection{RQ2: What Drives Reported Progress?}
\label{sec:results_progress_attribution}
%
As outlined in Section~\ref{sec:method_attribution}, we attribute progress by identifying contract field discrepancies between a system and its baselines. Figure~\ref{tab:progress-attribution} categorizes the primary drivers of reported improvements: protocol progress (22/62), workload or fidelity shifts (17/62), mixed progress (11/62), assumption tradeoffs (5/62), component progress (4/62), and hardware or deployment shifts (3/62).
A primary protocol-progress label marks the protocol change as the \emph{dominant} driver, not the only one: all 22 such papers carry at least one secondary non-protocol caveat.
While protocol progress contributes as a primary or secondary factor in 49 of 62 papers, zero papers achieve \emph{pure} protocol progress, a protocol-primary label with no non-protocol caveat. 
This is because every reported gain co-occurs with non-protocol factors such as workload choices, fidelity modifications, amortization, or reporting scope. 
This reflects pervasive entanglement rather than a lack of technical merit, demonstrating that performance claims (e.g., ``System A outperforms System B'') remain incomplete without specifying the underlying contract and source of gain.

\begin{tcolorbox}[
  enhanced,
  colback=blue!5,
  colframe=black,
  boxrule=0.5pt,
  arc=5pt,
  left=2pt, right=2pt, top=2pt, bottom=2pt,
  colbacktitle=gray!10,
  coltitle=black,
  fonttitle=\bfseries
]
{\bf Finding 2:} Cryptographic progress in PTI is documented (49 of 62 papers show it as a factor) but never isolated (0 of 62 show it alone), so every reported speedup is entangled with a change in model, workload, or accounting.
\end{tcolorbox}




\begin{figure}[t]
\centering
\includegraphics[width=\columnwidth]{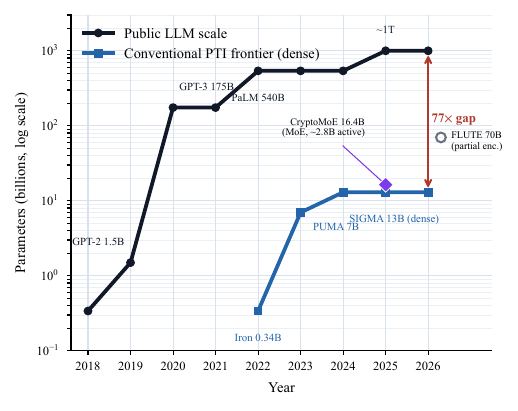}
\caption{Public LLM inference scale versus the audited conventional cryptographic
PTI frontier, on a logarithmic parameter axis. The dense frontier enters at 0.34B
(Iron, 2022), reaches 13B (SIGMA, 2024), and does not move through 2026, while
public models pass a trillion parameters, a gap of nearly two orders of magnitude.
CryptoMoE's 16.4B mixture-of-experts (about 2.8B active per token) sits off the
dense line, and FLUTE's 70B partial-encryption result is kept off the frontier as
a weaker contract. Each point carries its own contract
(Table~\ref{tab:scaling-contracts}), so the frontier is a scale-context upper
bound, not a like-for-like capability comparison.}
\label{fig:scaling-linear}
\end{figure}

\subsection{RQ3: The PTI Scaling Frontier}
\label{sec:results_scaling}
Section~\ref{sec:method_scaling} defines the annual PTI scaling frontier using the maximum model parameter count reported in PIC cards per year. As shown in Figure~\ref{fig:scaling-linear}, the conventional cryptographic frontier progressed from 0.34B parameters in 2022 to 7B in 2023 and 13B in 2024. 
However, the PTI frontier has since plateaued. 
Audited 2025--26 dense systems (7--8B) remain below SIGMA's 13B, while larger audited points require structural trade-offs, such as CryptoMoE's 16.4B mixture-of-experts or FLUTE's 70B model under partial encryption. Meanwhile, the unencrypted public baseline surged from GPT-3 (175B in 2020) and PaLM (540B in 2022) to $\sim$1T parameters by 2025. Thus, despite internal progress, PTI lags far behind public model deployment.

This private frontier represents an optimistic envelope compiled across heterogeneous evaluation contracts (Table~\ref{tab:scaling-contracts} in Appendix~\ref{sec:app-frontier-models}). 
Specifically, 2022's Iron relies on a hybrid 2PC setup, 2023's PUMA uses end-to-end 3PC, and 2024's SIGMA assumes an online-only protocol.
Therefore, this frontier merely tracks the largest scale demonstrated under any conventional cryptographic contract in a given year. 
Enforcing a single fixed contract (e.g., end-to-end 2PC) would lower the private frontier further, making the observed public--private scaling gap a conservative lower bound.

\begin{tcolorbox}[
  enhanced,
  colback=blue!5,
  colframe=black,
  boxrule=0.5pt,
  arc=5pt,
  left=2pt, right=2pt, top=2pt, bottom=2pt,
  colbacktitle=gray!10,
  coltitle=black,
  fonttitle=\bfseries
]
{\bf Finding 3:} The PTI frontier has plateaued at 13B since 2024 while the plaintext frontier has passed 1T parameters. Because the private curve generously aggregates non-comparable contracts, this gap is much wider in practice.
\end{tcolorbox}

\section{RQ4: The Cost of Deployment-Scale PTI}
\label{sec:cost_roadmap}

As discussed in  Section~\ref{sec:results_scaling}, the empirical PTI frontier remains bounded at 13B, while production LLMs have passed a trillion parameters.
Given this gap, what would it actually cost, in engineering terms, to run private inference at the scale of models now in production?
To quantify the operational cost of deployment-scale private inference, we extrapolate audited empirical metrics to 70B- and 405B-class model backbones, projecting per-token latency, communication volume, and offline preprocessing requirements.

\begin{table*}[t]
\centering
\footnotesize
\setlength{\tabcolsep}{3.5pt}
\renewcommand{\arraystretch}{1.15}
\caption{What one private forward pass (one generated token) would cost at deployment scale, extrapolated per family from published evidence under that family's own contract. 
The 3PC per-token anchors are the marginal
second token from the source's own table, so prefill is not considered.
}
\label{tab:price-at-scale}
\resizebox{\textwidth}{!}{%
\begin{tabular}{llllll}
\toprule
\textbf{Family (env.)} & \textbf{Contract caveat} & \textbf{Metric} & \textbf{7--13B (published)} & \textbf{70B (proj.)} & \textbf{405B (proj.)} \\
\midrule
FSS, 2PC+dealer (SIGMA) & online-only, GPU & time/pass & 23--38 s & $\sim$115 s & $\sim$384 s \\
 & & comm/pass & 12--19 GB & $\sim$58 GB & $\sim$188 GB \\
 & & \textbf{FSS key/query} & \textbf{255--419 GB} & \textbf{$\sim$1.4 TB} & \textbf{$\sim$4.9 TB} \\
\midrule
3PC secret sharing (PUMA) & end-to-end, 20 Gbps LAN & time/token & 164 s & $\sim$27 min & $\sim$2.6 h \\
 & & comm/token & 2.1 GB & $\sim$21 GB & $\sim$119 GB \\
\midrule
Hybrid HE+MPC, 2PC (BumbleBee) & end-to-end, CPU & time/token & 832 s & $\sim$2.3 h & $\sim$13 h \\
 & & comm/token & 5.6 GB & $\sim$18 GB & $\sim$43 GB \\
\midrule
HE non-interactive (NEXUS) & \textbf{batch-32 amortized}, 4$\times$A100 & time/token & 52 s & $\sim$7.6 min & $\sim$44 min \\
 & single-query (de-amortized) & time/token & $\sim$28 min & $\sim$4 h & $\sim$23 h \\
\bottomrule
\end{tabular}%
}
\end{table*}

{\bf Fitting a cost model to the corpus.} 
We construct a calibration dataset of 256 empirical performance tuples (latency, communication volume, rounds, and preprocessing overhead alongside network, hardware, and accounting contexts) spanning 20 systems from the audited corpus.
Measurements for each system are fitted to a power-law curve strictly within its isolated paper environment under a single evaluation contract.

When a paper reports cost at several model sizes (e.g., SIGMA spanning 0.124B to 13B parameters), we fit the power law $\mathrm{cost} \propto P^{\beta}$ to those points, where $\mathrm{cost}$ is the reported latency or communication, $P$ is parameter count, and $\beta$ is the fitted scaling exponent. Leave-one-out error stays below 12\% in every curve-fitting case.
When a paper reports cost at only one large model size, we scale that point using analytical computation exponents (e.g., linear layer compute scaling as $P$ and activation communication scaling as $\sqrt{P}$ at fixed sequence length). 
To evaluate sensitivity to this analytical exponent, we sweep $\beta$ across sub-linear to super-linear regimes ($\beta \in [0.8, 1.2]$) at 70B and 405B. 
The projected cost shifts by only a multiplicative factor across this range, so the order of magnitude of the projection matches the model scales.

{\bf The projected cost at production scale.}
Table~\ref{tab:price-at-scale} translates our fitted projections into per-token deployment costs. 
While plaintext inference for a 70B model on a single GPU node executes in tens of milliseconds per token, every private baseline exhibits a 3-to-5 order-of-magnitude overhead. This gap widens with scale as communication and preprocessing costs outpace compute. 
Evaluating the four protocol families, each under its own published contract (Table~\ref{tab:price-at-scale}), reveals a sharp divergence, ranging from roughly two minutes per token for the fastest online family to several hours for the slowest.
Appendix~\ref{app:cost-validation} stress-tests these projections against held-out model sizes and alternative scaling exponents; the order of magnitude holds in both cases.

{\bf Three structural walls.}
To identify the underlying drivers behind the stalled PTI scaling frontier, we examine the massive projected costs reported in Table~\ref{tab:price-at-scale}. 
We observe that PTI performance hits three \emph{structural walls}, representing fundamental physical and system limits that protocol adjustments alone cannot overcome. 
First, a \textbf{preprocessing wall} constrains FSS-based systems. 
While FSS offers low online latency, key sizes scale at $\beta{=}0.72$, requiring $\sim$1.4\,TB of offline dealer keys per 70B inference setup. Transporting this state requires $\sim$20 minutes over a 9.4\,Gbps LAN, dwarfing the 115-second online inference phase itself.
Second, a \textbf{bandwidth floor} is set by raw communication volume (18--58\,GB per token at 70B), imposing an absolute physical throughput limit (24--77 minutes per token on 100\,Mbps links) regardless of compute speed or round counts.
Third, an \textbf{amortization dependence} divides interactive and batched service contracts, as non-interactive HE reaches minute-scale 70B projections (about 7.6 minutes per token) only by batching 32 queries, while its single-query cost is roughly $30\times$ worse.

\begin{tcolorbox}[
  enhanced,
  colback=blue!5,
  colframe=black,
  boxrule=0.5pt,
  arc=5pt,
  left=2pt, right=2pt, top=2pt, bottom=2pt,
  colbacktitle=gray!10,
  coltitle=black,
  fonttitle=\bfseries
]
{\bf Finding 4:} Private inference at 70B costs three to five orders of magnitude more than plaintext, and that gap decomposes into three physical bottlenecks.
\end{tcolorbox}

\section{Discussion}
\label{sec:research_gaps}

\subsection{What This SoK Adds to PTI Progress}
\label{sec:discussion_progress}

Over just a few years, PTI has driven remarkable cryptographic momentum, scaling protocols from small encoders to billion-parameter decoders, slashing communication volumes by orders of magnitude, and establishing theoretical primitives (non-interactive HE at scale, FSS for non-linear activations, and dynamic MPC--HE hybrid routing) as practical. 
Sixty-two systems since 2022 reflect a rapidly evolving field producing novel algorithms and systems.
This SoK provides a progress lens to this rapidly evolving field. 
Because every paper reports under an isolated evaluation contract, the field has lacked the visibility to distinguish cumulative, composable advances from disconnected empirical wins. 
Our diagnostics on pairwise comparability (Section~\ref{sec:results_comparability}), progress attribution (Section~\ref{sec:results_progress_attribution}), and the PTI scaling frontier (Section~\ref{sec:results_scaling}) contextualize each gain in the literature and translate these diagnostics into concrete practice.

\subsection{Report the Contract, Not Just the Number}
\label{sec:discussion_contract}
Most articles already collect the core parameters required for their evaluation, including model parameters, workload regimes, privacy bounds, and cost metrics. 
What is missing is consolidating this data into a single, checkable reference. 
We recommend attaching a standardized \emph{PIC Card} (Table~\ref{tab:evaluation-contract-card}), covering scale, threat model, fidelity, deployment costs, and baselines. As Section~\ref{sec:results_comparability} demonstrates, assembling these fields at corpus scale establishes the precise specification required for future cross-paper comparisons.
A similar principle governs performance attribution. 
The contribution of a system is sharpened when we dissect speedups across explicit sources (protocol design, workload, relaxed security assumptions, hardware accounting, or component kernels as shown in Section~\ref{sec:method_attribution}). 
Transparent attribution enables future work to iterate on verified modular advances rather than inheriting unisolated performance numbers.

\subsection{Measure Scaling Against a Moving Frontier}
Section~\ref{sec:results_scaling} illustrates both the advance of the PTI frontier and the substantial distance remaining to plaintext models. 
While frontier model evaluation is not required for every system, authors should unambiguously scope their contribution across four progress settings: 
\textbf{(1) model-scale progress} reporting increasingly larger models under comparable privacy and cost bounds, \textbf{(2) cost-scaling progress} characterizing latency, communication, or rounds as workload dimensions grow, \textbf{(3) component-scaling progress} optimizing an isolated bottleneck operator without end-to-end claims, and \textbf{(4) deployment-scenario progress} justifying model scale and workload within a concrete application context. 
This suggested guideline provides a clear four-dimensional lens for evaluating system scaling.

\subsection{Benchmark with Deployment Scenarios}
Section~\ref{sec:results_comparability} illustrates the challenges of unstandardized evaluation.
MPC, HE, and hybrid systems, each engineered for distinct trade-offs, are flattened onto a single comparison. 
Hence, benchmarking should instead organize around explicit deployment scenarios so that systems compete on their intended settings:
 \textbf{(a) Interactive WAN} prioritizes round complexity and tail latency under realistic network bounds; \textbf{(b) Batch analytics} evaluates throughput and amortization, provided accounting is fully transparent; \textbf{(c) Collaborative inference} spotlights party topology and explicit leakage boundaries; \textbf{(d) Model-confidential inference} enforces equal protection for proprietary model weights and user inputs; and \textbf{(e) Component stress-testing} isolates bottleneck operators without conflating microbenchmarks with full-system feasibility.

 \subsection{Matching Backbones to Deployments}
\label{sec:discussion_backbone_choice}
A practical question follows: when facing a real-world deployment, which cryptographic backbone should a practitioner choose? Because each cryptographic family exhibits a distinct cost profile, optimal selection requires mapping these cost structures to a deployment's primary bottleneck (Table~\ref{tab:backbone-guidance}, Appendix~\ref{app:extra_tables}).
We categorize the deployment options as follows: 
\textbf{Interactive WAN}: Low-round HE or few-round hybrids (minimizing round-trip latency overhead);
\textbf{Batch throughput}: SIMD-packed HE or amortized-offline hybrids (maximizing parallel efficiency);
\textbf{Low-latency LAN}: MPC or hybrid schemes (leveraging cheap, highly interactive rounds);
\textbf{Bandwidth-constrained links}: HE or communication-optimized protocols (minimizing data transfer);
\textbf{Model confidentiality}: Dual input-and-model 2PC (ensuring weight privacy); and
\textbf{Helper-assisted settings}: 3PC or server-aided MPC (when a non-colluding party is available).

These mappings reflect architectural trade-offs rather than absolute rules. For instance, NEXUS leverages a single-round design to minimize WAN latency, but pays for it in heavy homomorphic computation. 
Conversely, round-intensive MPC frameworks achieve exceptional LAN throughput that degrades severely over a WAN. Hence, selecting a cryptographic backbone requires first formalizing the deployment settings, not the other way around.

\section{Limitations}
\label{sec:limitations}

This SoK audits published evidence rather than reproducing performance on unified hardware. 
Hence, all findings in the paper evaluate what the published record actually supports.
We summarize some limitations of our approach.

{\bf Corpus scope.} We strictly evaluate conventional cryptographic PTI, excluding split-inference and statistical approaches. Scaling findings apply explicitly to cryptographic setups rather than all private inference frameworks.

{\bf Contract labeling.} Labeling relies on conservative defaults (e.g., preferring unknown over inferred), backed by blind re-extraction and sensitivity analysis (Appendix~\ref{sec:results_sensitivity}). The corpus maps empirical evidence rather than ranking systems.

{\bf Gate granularity.} Gate compliance certifies that a pair is \emph{not provably incomparable} rather than identical. Coarse categorical fields (e.g., fidelity approximations, network classes, and cost accounting bounds) can obscure subtle reporting discrepancies. For instance, while Euston and NEXUS clear every gate as a deliberate reimplementation, the two papers dispute whether user-side preprocessing belongs in end-to-end latency.

{\bf Scale as a capability proxy.} We measure the PTI frontier using parameter count. Although parameter count omits architectural quality and context length, it directly dictates cryptographic compute costs, making it highly relevant for our scaling analysis.

\section{Conclusions}
This SoK asks whether PTI's reported progress accumulates toward deployment-scale private inference or merely produces isolated points that cannot be assembled into a trajectory. 
Our four analyses point to the same answer from distinct perspectives. 
First, the scaling frontier shows that demonstrated capability has plateaued well below production scale, and that even this frontier is an optimistic upper bound stitched from mutually incompatible contracts. 
Second, the comparability audit explains {\em why}: virtually no two published claims share enough evaluation context to be directly compared, with aligning pairs clustering around a few hub systems rather than spanning the corpus. 
Third, the attribution analysis reveals {\em what} this fragmentation obscures: while cryptographic progress is widespread, it is never the sole driver of reported gains, meaning a headline speedup does not inherently signify protocol advancement. 
Finally, the cost projection shows {\em how} to close this gap: it is not a single metric, but overcoming three distinct structural walls, each demanding a dedicated resolution.

Taken together, these results reframe what constitutes valid evidence of progress in this field. 
A faster system is uninformative in isolation unless it explicitly specifies its scale, governing contract, and exact source of performance gain. 
Individual PTI studies have delivered remarkable technical advances over the years.
What the field needs next is the reporting discipline to show that the speedups are converging on deployment.

\cleardoublepage
\section*{Ethical Considerations}
This paper systematizes already-published research and raises no ethical concerns of its own. It studies only the public record: peer-reviewed papers, their publicly released artifacts, and public model-scale references. It involves no human subjects, no personal or sensitive data, and introduces no attacks, exploits, or deanonymization techniques. We name systems solely to analyze the evaluation contracts behind their published results; our comparability and attribution labels concern what the published evidence can support, not the merit or conduct of any author. The only automated processing in our pipeline is an LLM verifier that re-reads the same published text as a second, blind check on the human extraction; it processes no data beyond the papers themselves, and its outputs are released with the artifacts. We therefore see no risk of harm arising from this work.

\section*{Open Science}
In keeping with the open science policy, we release the artifacts behind this
systematization: the frozen 62-paper corpus metadata, the structured extraction
protocol and codebook (reproduced in Appendix~\ref{app:llm_protocol}), the
per-paper evaluation-contract fields, the comparability-gate implementation and
the pairwise graph it produces, the progress-attribution labels, the blind
re-extraction records, the cost-model calibration data, and the scripts that
regenerate every table and figure from these inputs. Together they let a reader
reproduce each reported number directly from the released data. We do not
redistribute the audited papers or their full text, which remain under their
publishers' copyright; the corpus references each work by citation and releases
only derived metadata and the short evidence quotes needed to audit each field.

The artifacts are archived at \url{https://doi.org/10.5281/zenodo.22105363}.
The record is currently deposited anonymously so that it remains usable
alongside double-blind submissions of this work; we will publish a
de-anonymized version of the same record, under the same concept DOI, once the
paper is accepted.

\bibliographystyle{IEEEtran}
\bibliography{files/refs}

@article{pindado2026sok,
  title={SoK of Private Deep Neural Network Inference with Approximate Fully Homomorphic Encryption},
  author={Pindado, Zaira and Spendlhofer, Thomas and Allam, Mohamed and Mehta, Priyam and Martens, Lena and Pe{\~n}a, Antonio J},
  journal={Cryptology ePrint Archive},
  year={2026}
}

@article{pope2023efficiently,
  title={Efficiently scaling transformer inference},
  author={Pope, Reiner and Douglas, Sholto and Chowdhery, Aakanksha and Devlin, Jacob and Bradbury, James and Heek, Jonathan and Xiao, Kefan and Agrawal, Shivani and Dean, Jeff},
  journal={Proceedings of machine learning and systems},
  volume={5},
  pages={606--624},
  year={2023}
}

@article{li2024survey,
  title={A Survey on Private Transformer Inference},
  author={Li, Yang and Zhou, Xinyu and Wang, Yitong and Qian, Liangxin and Zhao, Jun},
  journal={arXiv preprint arXiv:2412.08145},
  year={2024}
}

@article{li2025survey,
  title={Private Transformer Inference in {MLaaS: A} Survey},
  author={Li, Yang and Zhou, Xinyu and Wang, Yitong and Qian, Liangxin and Zhao, Jun},
  journal={arXiv preprint arXiv:2505.10315},
  year={2025}
}

@inproceedings{ng2023sok,
  title={Sok: Cryptographic neural-network computation},
  author={Ng, Lucien KL and Chow, Sherman SM},
  booktitle={2023 IEEE Symposium on Security and Privacy (SP)},
  pages={497--514},
  year={2023},
  organization={IEEE}
}

@article{gouert2023sok,
  title={Sok: New insights into fully homomorphic encryption libraries via standardized benchmarks},
  author={Gouert, Charles and Mouris, Dimitris and Tsoutsos, Nektarios},
  journal={Proceedings on Privacy Enhancing Technologies},
  volume={2023},
  number={3},
  pages={154--172},
  year={2023}
}

@article{falcetta2022he,
  title={Privacy-preserving deep learning with homomorphic encryption: An introduction},
  author={Falcetta, Alessandro and Roveri, Manuel},
  journal={IEEE Computational Intelligence Magazine},
  volume={17},
  number={3},
  pages={14--25},
  year={2022},
  publisher={IEEE}
}

@article{podschwadt2021sok,
  title={Sok: Privacy-preserving deep learning with homomorphic encryption},
  author={Podschwadt, Robert and Takabi, Daniel and Hu, Peizhao},
  journal={arXiv preprint arXiv:2112.12855},
  year={2021}
}

@misc{xue2025sokfullyhomomorphicencryption,
      title={SoK: Can Fully Homomorphic Encryption Support General AI Computation? A Functional and Cost Analysis}, 
      author={Jiaqi Xue and Xin Xin and Wei Zhang and Mengxin Zheng and Qianqian Song and Minxuan Zhou and Yushun Dong and Dongjie Wang and Xun Chen and Jiafeng Xie and Liqiang Wang and David Mohaisen and Hongyi Wu and Qian Lou},
      year={2025},
      eprint={2504.11604},
      archivePrefix={arXiv},
      primaryClass={cs.CR},
      url={https://arxiv.org/abs/2504.11604}, 
}

@inproceedings{yan2024protecting,
  title={On protecting the data privacy of large language models (LLMs): A survey},
  author={Yan, Biwei and Li, Kun and Xu, Minghui and Dong, Yueyan and Zhang, Yue and Ren, Zhaochun and Cheng, Xiuzhen},
  booktitle={2024 International Conference on Meta Computing (ICMC)},
  pages={1--12},
  year={2024},
  organization={IEEE}
}

@inproceedings{bumblebee2023,
  title={Bumblebee: Secure two-party inference framework for large transformers},
  author={Lu, Wen-jie and Huang, Zhicong and Gu, Zhen and Li, Jingyu and Liu, Jian and Hong, Cheng and Ren, Kui and Wei, Tao and Chen, WenGuang},
  booktitle={Network and Distributed System Security Symposium, NDSS},
  volume={2025},
  year={2025}
}

@inproceedings{shaft2025,
  title={SHAFT: Secure, Handy, Accurate, and Fast Transformer Inference},
  author={Kei, Andes YL and Chow, Sherman SM},
  booktitle={Network and Distributed System Security Symposium, NDSS},
  volume={2025},
  year={2025}
}

@inproceedings{blb2025,
  title={Breaking the layer barrier: Remodeling private transformer inference with hybrid $\{$CKKS$\}$ and $\{$MPC$\}$},
  author={Xu, Tianshi and Lu, Wen-jie and Yu, Jiangrui and Chen, Yi and Lin, Chenqi and Wang, Runsheng and Li, Meng},
  booktitle={34th USENIX Security Symposium (USENIX Security 25)},
  pages={2653--2672},
  year={2025}
}

@article{sigma2023,
  title={Sigma: Secure gpt inference with function secret sharing},
  author={Gupta, Kanav and Jawalkar, Neha and Mukherjee, Ananta and Chandran, Nishanth and Gupta, Divya and Panwar, Ashish and Sharma, Rahul},
  journal={Proceedings on Privacy Enhancing Technologies},
  volume={2024},
  number={4},
  pages={61--79},
  year={2024}
}

@inproceedings{mpcformer2023,
  title={Mpcformer: fast, performant and private transformer inference with mpc},
  author={Li, Dacheng and Shao, Rulin and Wang, Hongyi and Guo, Han and Xing, Eric P and Zhang, Hao},
  booktitle={The Eleventh International Conference on Learning Representations (ICLR)},
  year={2023}
}

@article{puma2023,
  title={Puma: Secure inference of llama-7b in five minutes},
  author={Dong, Ye and Lu, Wen-jie and Zheng, Yancheng and Wu, Haoqi and Zhao, Derun and Tan, Jin and Huang, Zhicong and Hong, Cheng and Wei, Tao and Chen, Wenguang},
  journal={arXiv preprint arXiv:2307.12533},
  year={2023}
}

@inproceedings{secformer2024,
  title={Secformer: fast and accurate privacy-preserving inference for transformer models via SMPC},
  author={Luo, Jinglong and Zhang, Yehong and Zhang, Zhuo and Zhang, Jiaqi and Mu, Xin and Wang, Hui and Yu, Yue and Xu, Zenglin},
  booktitle={Findings of the Association for Computational Linguistics ACL 2024},
  pages={13333--13348},
  year={2024}
}

@inproceedings{privformer2023,
  title={Privformer: Privacy-preserving transformer with mpc},
  author={Akimoto, Yoshimasa and Fukuchi, Kazuto and Akimoto, Youhei and Sakuma, Jun},
  booktitle={2023 IEEE 8th European Symposium on Security and Privacy (EuroS\&P)},
  pages={392--410},
  year={2023},
  organization={IEEE}
}

@inproceedings{ditto2024,
  title={Ditto: Quantization-aware secure inference of transformers upon mpc},
  author={Wu, Haoqi and Fang, Wenjing and Zheng, Yancheng and Ma, Junming and Tan, Jin and Wang, Lei},
  booktitle={International Conference on Machine Learning (ICML)},
  pages={53346--53365},
  year={2024}
}

@inproceedings{nexus2024,
  title={Secure transformer inference made non-interactive},
  author={Zhang, Jiawen and Yang, Xinpeng and He, Lipeng and Chen, Kejia and Lu, Wen-jie and Wang, Yinghao and Hou, Xiaoyang and Liu, Jian and Ren, Kui and Yang, Xiaohu},
  booktitle={Network and Distributed System Security Symposium, NDSS},
  volume={2025},
  year={2025}
}

@inproceedings{bolt2024,
  title={Bolt: Privacy-preserving, accurate and efficient inference for transformers},
  author={Pang, Qi and Zhu, Jinhao and M{\"o}llering, Helen and Zheng, Wenting and Schneider, Thomas},
  booktitle={2024 IEEE Symposium on Security and Privacy (SP)},
  pages={4753--4771},
  year={2024},
  organization={IEEE}
}

@inproceedings{cipherprune2025,
  title={Cipherprune: Efficient and scalable private transformer inference},
  author={Zhang, Yancheng and Xue, Jiaqi and Zheng, Mengxin and Xie, Mimi and Zhang, Mingzhe and Jiang, Lei and Lou, Qian},
  booktitle={The Thirteenth International Conference on Learning Representations (ICLR)},
  year={2025}
}

@article{ciphergpt2023,
  title={Ciphergpt: Secure two-party gpt inference},
  author={Hou, Xiaoyang and Liu, Jian and Li, Jingyu and Li, Yuhan and Lu, Wen-jie and Hong, Cheng and Ren, Kui},
  journal={Cryptology ePrint Archive},
  year={2023}
}

@article{pptif2024,
  title={PPTIF: Privacy-preserving transformer inference framework for language translation},
  author={Liu, Yanxin and Su, Qianqian},
  journal={IEEE Access},
  volume={12},
  pages={48881--48897},
  year={2024},
  publisher={IEEE}
}

@article{privtf2025,
  title={PrivTF: Efficient and Privacy-Preserving Transformer Inference With Optimized Protocols for Secure NLP},
  author={Peng, Ziwei and Xue, Kaiping and Zhu, Bin and Huang, Yaxuan and Li, Jian and Hao, Meng and Zhang, Tianwei},
  journal={IEEE Transactions on Network Science and Engineering},
  year={2025},
  publisher={IEEE}
}

@article{iron2022,
  title={Iron: Private inference on transformers},
  author={Hao, Meng and Li, Hongwei and Chen, Hanxiao and Xing, Pengzhi and Xu, Guowen and Zhang, Tianwei},
  journal={Advances in neural information processing systems},
  volume={35},
  pages={15718--15731},
  year={2022}
}

@inproceedings{thex2022,
  title={The-x: Privacy-preserving transformer inference with homomorphic encryption},
  author={Chen, Tianyu and Bao, Hangbo and Huang, Shaohan and Dong, Li and Jiao, Binxing and Jiang, Daxin and Zhou, Haoyi and Li, Jianxin and Wei, Furu},
  booktitle={Findings of the Association for Computational Linguistics: ACL 2022},
  pages={3510--3520},
  year={2022}
}

@article{nimbus2024,
  title={Nimbus: Secure and efficient two-party inference for transformers},
  author={Li, Zhengyi and Yang, Kang and Tan, Jin and Lu, Wen-jie and Wu, Haoqi and Wang, Xiao and Yu, Yu and Zhao, Derun and Zheng, Yancheng and Guo, Minyi and others},
  journal={Advances in Neural Information Processing Systems},
  volume={37},
  pages={21572--21600},
  year={2024}
}

@article{securetlm2024,
  title={SecureTLM: Private inference for transformer-based large model with MPC},
  author={Chen, Yuntian and Meng, Xianjia and Shi, Zhiying and Ning, Zhiyuan and Lin, Jingzhi},
  journal={Information Sciences},
  volume={667},
  pages={120429},
  year={2024},
  publisher={Elsevier}
}

@article{marill2024,
  title={Mpc-minimized secure llm inference},
  author={Rathee, Deevashwer and Li, Dacheng and Stoica, Ion and Zhang, Hao and Popa, Raluca},
  journal={arXiv preprint arXiv:2408.03561},
  year={2024}
}

@article{cascade2025,
  title={Cascade: token-sharded private LLM inference},
  author={Thomas, Rahul and Zahran, Louai and Choi, Erica and Potti, Akilesh and Goldblum, Micah and Pal, Arka},
  journal={arXiv preprint arXiv:2507.05228},
  year={2025}
}

@inproceedings{cipherformer2024,
  title={CipherFormer: Efficient Transformer Private Inference with Low Round Complexity},
  author={Wang, Weize and Kuang, Yi},
  booktitle={2024 27th International Conference on Computer Supported Cooperative Work in Design (CSCWD)},
  pages={3054--3059},
  year={2024},
  organization={IEEE}
}

@article{east2025,
  title={East: Efficient and Accurate Secure Inference Framework for Transformer},
  author={Ding, Yuanchao and Guo, Hua and Guan, Yewei and Liu, Weixin and Huo, Jiarong and Guan, Zhenyu and Zhang, Xiyong},
  journal={IEEE Transactions on Services Computing},
  year={2025},
  publisher={IEEE}
}

@article{practicalllm2025,
  title={Practical Secure Inference Algorithm for Fine-tuned Large Language Model Based on Fully Homomorphic Encryption},
  author={Zhang, Ruoyan and Zheng, Zhongxiang and Bao, Wankang},
  journal={arXiv preprint arXiv:2501.01672},
  year={2025}
}

@inproceedings{encryptedllm2025,
  title={EncryptedLLM: Privacy-preserving large language model inference via GPU-accelerated fully homomorphic encryption},
  author={De Castro, Leo and Escudero, Daniel and Agrawal, Adya and Polychroniadou, Antigoni and Veloso, Manuela},
  booktitle={Forty-second International Conference on Machine Learning},
  year={2025}
}

@article{ensi2025,
  title={ENSI: Efficient Non-Interactive Secure Inference for Large Language Models},
  author={He, Zhiyu and Wang, Maojiang and Gao, Xinwen and Luo, Yuchuan and Liu, Lin and Fu, Shaojing},
  journal={arXiv preprint arXiv:2509.09424},
  year={2025}
}

@article{secbert2024,
  title={SecBERT: Privacy-preserving pre-training based neural network inference system},
  author={Huang, Hai and Wang, Yongjian},
  journal={Neural Networks},
  volume={172},
  pages={106135},
  year={2024},
  publisher={Elsevier}
}

@article{securegpt2024,
  title={SecureGPT: A Framework for Multi-Party Privacy-Preserving Transformer Inference in GPT},
  author={Zeng, Chenkai and He, Debiao and Feng, Qi and Yang, Xiaolin and Luo, Qingcai},
  journal={IEEE Transactions on Information Forensics and Security},
  year={2024},
  publisher={IEEE}
}

@article{permllm2024,
  title={Permllm: Private inference of large language models within 3 seconds under wan},
  author={Zheng, Fei and Chen, Chaochao and Han, Zhongxuan and Zheng, Xiaolin},
  journal={arXiv preprint arXiv:2405.18744},
  year={2024}
}

@inproceedings{privmllm2025,
  title={PrivMLLM: Efficient Three-Party Multimodal Large Language Model Secure Inference Supported Prompt Privacy},
  author={Xue, Yujie and Liu, Lin and Luo, Yuchuan and Sun, Bing and Fu, Shaojing},
  booktitle={2025 IEEE 33rd International Conference on Network Protocols (ICNP)},
  pages={1--24},
  year={2025},
  organization={IEEE}
}

@article{stip2023,
  title={Secure Transformer Inference Protocol},
  author={Yuan, Mu and Zhang, Lan and Li, Xiang-Yang},
  journal={arXiv preprint arXiv:2312.00025},
  year={2023}
}

@inproceedings{hedecoding2025,
  title={Efficient Decoding Methods for Language Models on Encrypted Data},
  author={Avitan, Matan and Baruch, Moran and Drucker, Nir and Zimerman, Itamar and Goldberg, Yoav},
  booktitle={Findings of the Association for Computational Linguistics: IJCNLP-AACL 2025},
  year={2025}
}

@inproceedings{primer2023,
  title={Primer: Fast private transformer inference on encrypted data},
  author={Zheng, Mengxin and Lou, Qian and Jiang, Lei},
  booktitle={2023 60th ACM/IEEE Design Automation Conference (DAC)},
  pages={1--6},
  year={2023},
  organization={IEEE}
}

@inproceedings{merge2024,
  title={Merge: Fast private text generation},
  author={Liang, Zi and Wang, Pinghui and Zhang, Ruofei and Xu, Nuo and Zhang, Shuo and Xing, Lifeng and Bai, Haitao and Zhou, Ziyang},
  booktitle={Proceedings of the AAAI Conference on Artificial Intelligence},
  volume={38},
  number={18},
  pages={19884--19892},
  year={2024}
}

@inproceedings{comet2025,
  title={Comet: Accelerating Private Inference for Large Language Model by Predicting Activation Sparsity},
  author={Yan, Guang and Zhang, Yuhui and Guo, Zimu and Zhao, Lutan and Chen, Xiaojun and Wang, Chen and Wang, Wenhao and Meng, Dan and Hou, Rui},
  booktitle={2025 IEEE Symposium on Security and Privacy (SP)},
  pages={2827--2845},
  year={2025},
  organization={IEEE}
}

@inproceedings{rhombus2024,
  title={Rhombus: Fast homomorphic matrix-vector multiplication for secure two-party inference},
  author={He, Jiaxing and Yang, Kang and Tang, Guofeng and Huang, Zhangjie and Lin, Li and Wei, Changzheng and Yan, Ying and Wang, Wei},
  booktitle={Proceedings of the 2024 on ACM SIGSAC Conference on Computer and Communications Security},
  pages={2490--2504},
  year={2024}
}

@article{powersoftmax2024,
  title={Power-softmax: Towards secure llm inference over encrypted data},
  author={Zimerman, Itamar and Adir, Allon and Aharoni, Ehud and Avitan, Matan and Baruch, Moran and Drucker, Nir and Lerner, Jenny and Masalha, Ramy and Meiri, Reut and Soceanu, Omri},
  journal={arXiv preprint arXiv:2410.09457},
  year={2024}
}

@inproceedings{encryptionfriendlyllm2024,
  title={Encryption-friendly LLM architecture},
  author={Rho, Donghwan and Kim, Taeseong and Park, Minje and Kim, Jung Woo and Chae, Hyunsik and Ryu, Ernest K and Cheon, Jung Hee},
  booktitle={The Thirteenth International Conference on Learning Representations (ICLR)},
  year={2025}
}

@article{llmencryptedprompt2023,
  title={Llms can understand encrypted prompt: Towards privacy-computing friendly transformers},
  author={Liu, Xuanqi and Liu, Zhuotao},
  journal={arXiv preprint arXiv:2305.18396},
  year={2023}
}

@article{vaswani2017attention,
  title={Attention is all you need},
  author={Vaswani, Ashish and Shazeer, Noam and Parmar, Niki and Uszkoreit, Jakob and Jones, Llion and Gomez, Aidan N and Kaiser, {\L}ukasz and Polosukhin, Illia},
  journal={Advances in neural information processing systems},
  volume={30},
  year={2017}
}

@inproceedings{mpcvit2023,
  title={Mpcvit: Searching for accurate and efficient mpc-friendly vision transformer with heterogeneous attention},
  author={Zeng, Wenxuan and Li, Meng and Xiong, Wenjie and Tong, Tong and Lu, Wen-jie and Tan, Jin and Wang, Runsheng and Huang, Ru},
  booktitle={Proceedings of the IEEE/CVF International Conference on Computer Vision},
  pages={5052--5063},
  year={2023}
}

@inproceedings{salvit2023,
  title={Sal-vit: Towards latency efficient private inference on vit using selective attention search with a learnable softmax approximation},
  author={Zhang, Yuke and Chen, Dake and Kundu, Souvik and Li, Chenghao and Beerel, Peter A},
  booktitle={Proceedings of the IEEE/CVF International Conference on Computer Vision},
  pages={5116--5125},
  year={2023}
}

@inproceedings{rnavit2023,
  title={RNA-ViT: reduced-dimension approximate normalized attention vision transformers for latency efficient private inference},
  author={Chen, Dake and Zhang, Yuke and Kundu, Souvik and Li, Chenghao and Beerel, Peter A},
  booktitle={2023 IEEE/ACM International Conference on Computer Aided Design (ICCAD)},
  pages={1--9},
  year={2023},
  organization={IEEE}
}

@article{aero2025,
  title={Entropy-Guided Attention for Private LLMs},
  author={Jha, Nandan Kumar and Reagen, Brandon},
  journal={arXiv preprint arXiv:2501.03489},
  year={2025}
}

@inproceedings{curl2024,
  title={Curl: Private LLMs through wavelet-encoded look-up tables},
  author={Santos, Manuel B and Mouris, Dimitris and Ugurbil, Mehmet and Jarecki, Stanislaw and Reis, Jos{\'e} and Sengupta, Shubho and de Vega, Miguel},
  booktitle={Conference on Applied Machine Learning in Information Security (CAMLIS)},
  year={2024}
}

@article{al2026sok,
  title={SoK: Private LLM Inference using Approximate Homomorphic Encryption},
  author={Al Badawi, Ahmad and Alexandru, Andreea and Polyakov, Yuriy and Vaikuntanathan, Vinod},
  journal={Cryptology ePrint Archive},
  year={2026}
}

@article{cryptogen2026,
  title={CryptoGen: Secure Transformer Generation with Encrypted KV-Cache Reuse},
  author={Zhang, Hedong and Javidnia, Neusha and Pardeshi, Shweta and Lou, Qian and Koushanfar, Farinaz},
  journal={arXiv preprint arXiv:2602.08798},
  year={2026}
}

@article{cachemir2026,
  title={Cachemir: Fully Homomorphic Encrypted Inference of Generative Large Language Model with KV Cache},
  author={Yu, Ye and Zhou, Yifan and Chen, Yi and Soto, Pedro and Xiong, Wenjie and Li, Meng},
  journal={arXiv preprint arXiv:2602.11470},
  year={2026}
}

@inproceedings{secmoe2026,
  title={SecMoE: Communication-Efficient Secure MoE Inference via Select-Then-Compute},
  author={Shen, Bowen and Chen, Yuyue and Yang, Peng and Zhang, Bin and Zhang, Xi and Jiang, Zoe L.},
  booktitle={Proceedings of the AAAI Conference on Artificial Intelligence},
  year={2026}
}

@article{aegis2026,
  title={AEGIS: Scaling Long-Sequence Homomorphic Encrypted Transformer Inference via Hybrid Parallelism on Multi-GPU Systems},
  author={Gong, Zhaoting and Ran, Ran and Yao, Fan and Wen, Wujie},
  journal={arXiv preprint arXiv:2604.03425},
  year={2026}
}

@article{encformer2026,
  title={EncFormer: Secure and Efficient Transformer Inference over Encrypted Data},
  author={Zhu, Yufan and Jin, Chao and Aung, Khin Mi Mi and Xiao, Xiaokui},
  journal={arXiv preprint arXiv:2604.09975},
  year={2026}
}

@inproceedings{moon2025thor,
  title={THOR: Secure transformer inference with homomorphic encryption},
  author={Moon, Jungho and Yoo, Dongwoo and Jiang, Xiaoqian and Kim, Miran},
  booktitle={Proceedings of the 2025 ACM SIGSAC Conference on Computer and Communications Security},
  pages={3765--3779},
  year={2025}
}

@inproceedings{zhang2025moai,
  title={MOAI: Module-optimizing architecture for non-interactive secure transformer inference},
  author={Zhang, Linru and Wang, Xiangning and Sim, Jun Jie and Huang, Zhicong and Zhong, Jiahao and Wang, Huaxiong and Duan, Pu and Lam, Kwok-Yan},
  booktitle={The Fourteenth International Conference on Learning Representations},
  year={2026}
}

@misc{epochai,
  author={{Epoch AI}},
  title={Data on Notable {AI} Models},
  year={2026},
  howpublished={\url{https://epoch.ai/data/ai-models}},
  note={Accessed August 2026}
}

@inproceedings{park2025powerformer,
  title={Powerformer: Efficient and High-Accuracy Privacy-Preserving Language Model with Homomorphic Encryption},
  author={Park, Dongjin and Lee, Eunsang and Lee, Joon-Woo},
  booktitle={Annual Meeting of the Association for Computational Linguistics (ACL)},
  year={2025}
}

@article{yubeaton2024truncformer,
  title={TruncFormer: Private LLM Inference Using Only Truncations},
  author={Yubeaton, Patrick and Mo, Jianqiao Cambridge and Garimella, Karthik and Jha, Nandan Kumar and Reagen, Brandon and Hegde, Chinmay and Garg, Siddharth},
  journal={arXiv preprint arXiv:2412.01042},
  year={2024}
}

@inproceedings{li2025privategpt,
  title={An Efficient Private {GPT} Never Autoregressively Decodes},
  author={Li, Zhengyi and Guan, Yue and Yang, Kang and Feng, Yu and Liu, Ning and Yu, Yu and Leng, Jingwen and Guo, Minyi},
  booktitle={International Conference on Machine Learning (ICML)},
  year={2025}
}

@article{zeng2025ppmlreview,
  title={Towards Efficient Privacy-Preserving Machine Learning: A Systematic Review from Protocol, Model, and System Perspectives},
  author={Zeng, Wenxuan and Xu, Tianshi and Chen, Yi and Zhou, Yifan and Zhang, Mingzhe and Tan, Jin and Hong, Cheng and Li, Meng},
  journal={arXiv preprint arXiv:2507.14519},
  year={2025}
}

@inproceedings{li2026shuffling,
  title={On the (In-)Security of the Shuffling Defense in the Transformer Secure Inference},
  author={Li, Zhengyi and Wang, Yakai and Yang, Kang and Yu, Yu and Gui, Jiaping and Feng, Yu and Liu, Ning and Guo, Minyi and Leng, Jingwen},
  booktitle={Annual Meeting of the Association for Computational Linguistics (ACL)},
  year={2026}
}

@inproceedings{fan2026splitinference,
  title={What Does the Server See? Understanding Privacy Leakage from Large Language Models in Split Inference},
  author={Fan, Mingyuan and Liu, Yu and Wang, Fuyi and Chen, Cen},
  booktitle={Proceedings of the 2026 ACM SIGSAC Conference on Computer and Communications Security (CCS)},
  year={2026}
}

@article{park2026scaling,
  title={Scaling up {FHE}-based Privacy-Preserving Machine Learning: Higher Throughput, Longer Inputs for {LLaMA-3-8B}},
  author={Park, Jaiyoung and Park, Sejin and Park, Jai Hyun and Ahn, Jung Ho and Cheon, Jung Hee and Hanrot, Guillaume and Kim, Jung Woo and Park, Minje and Stehl{\'e}, Damien},
  journal={arXiv preprint arXiv:2601.18511},
  year={2026}
}

@inproceedings{chen2026sok,
  title = {SoK: Private Transformer-Based Model Inference},
  author = {Chen, Yuntian and Lu, Tianpei and Tang, Zhanyong and Zhang, Bingsheng and Shi, Zhiying and Luan, Yuxiang and Wang, Zhuzhu},
  booktitle = {35th USENIX Security Symposium (USENIX Security 26)},
  year = {2026}
}

@inproceedings{euston2026,
  title={Euston: Efficient and User-Friendly Secure Transformer Inference with Non-Interactivity},
  author={Gao, Xinwen and Fu, Shaojing and Liu, Lin and Liu, Zhuotao and Luo, Yuchuan and Wang, Yongjun},
  booktitle={2026 IEEE Symposium on Security and Privacy (SP)},
  year={2026}
}

@article{flute2026,
  title={{FLUTE}: {FSS}-Based Secure Two-Party {LLM} Inference Using Partial Transformer Encryption},
  author={Xue, Yujie and Liu, Lin and Luo, Yuchuan and Sun, Bing and Fu, Shaojing},
  journal={IEEE Transactions on Dependable and Secure Computing},
  volume={23},
  number={1},
  year={2026}
}

@article{atlas2026,
  title={{ATLAS}: Automated Approximation of Transformers for Efficient Homomorphic Inference in One Hour},
  author={Xie, Jianhang and Tan, Sicheng and Boddeti, Vishnu Naresh and Lu, Zhichao},
  journal={arXiv preprint arXiv:2607.23478},
  year={2026}
}

@inproceedings{smash2026,
  title={{SMASH}: Scalable Maliciously Secure Hybrid Multi-party Computation Framework for Privacy-Preserving Large Language Models},
  author={Lv, Yunlv and Zhang, Rui and Zhang, Zhiyuan and Wan, Ziyi and Zhang, Lanxue and Xue, Minhui and Li, Jiangtao and Cao, Yanan},
  booktitle={35th USENIX Security Symposium (USENIX Security 26)},
  year={2026}
}

@inproceedings{cryptomoe2025,
  title={{CryptoMoE}: Privacy-Preserving and Scalable Mixture of Experts Inference via Balanced Expert Routing},
  author={Zhou, Yifan and Xu, Tianshi and Hong, Jue and Wu, Ye and Li, Meng},
  booktitle={Advances in Neural Information Processing Systems (NeurIPS)},
  year={2025}
}

@inproceedings{mosformer2025,
  title={{Mosformer}: Maliciously Secure Three-Party Inference Framework for Large Transformers},
  author={Cheng, Ke and Xia, Yuheng and Song, Anxiao and Fu, Jiaxuan and Qu, Wenjie and Shen, Yulong and Zhang, Jiaheng},
  booktitle={Proceedings of the 2025 ACM SIGSAC Conference on Computer and Communications Security (CCS)},
  year={2025}
}

@article{fusefss2026,
  title={{FuseFSS}: Efficient Secure {LLM} Inference with Function Secret Sharing},
  author={Ma, Yuhan and Li, Yong and Schmid, Stefan},
  journal={arXiv preprint arXiv:2606.09551},
  year={2026}
}

@inproceedings{mpcache2025,
  title={{MPCache}: {MPC}-Friendly {KV} Cache Eviction for Efficient Private {LLM} Inference},
  author={Zeng, Wenxuan and Dong, Ye and Zhou, Jinjin and Tan, Jin and Wang, Lei and Wei, Tao and Wang, Runsheng and Li, Meng},
  booktitle={Advances in Neural Information Processing Systems (NeurIPS)},
  year={2025}
}

@misc{ellmo2026,
  title={{ELLMo}: Packing- and Depth-Aware Encrypted Transformer Inference},
  author={Guzelhan, Seyda Nur and Daksha, Lohit and Agull{\'o} Domingo, Carlos and Jonatan, Gilbert and Kim, John and Abell{\'a}n, Jos{\'e} L. and Kaeli, David and Joshi, Ajay},
  howpublished={Cryptology ePrint Archive, Paper 2026/198},
  year={2026}
}

@inproceedings{pritran2026,
  title={{PriTran}: Privacy-Preserving Inference for Transformer-Based Language Models under Fully Homomorphic Encryption},
  author={Mu, Yuechen and Li, Guangli and Chen, Shiping and others},
  booktitle={2026 IEEE/ACM International Symposium on Code Generation and Optimization (CGO)},
  year={2026}
}
\nocite{*}

\clearpage
\appendices
\section{Additional Result Tables}
\label{app:extra_tables}

This appendix collects the supporting result tables: the reporting-sufficiency
distribution over the 62-paper corpus and the full sensitivity analysis behind
the robustness checks.


\begin{table}[b]
\centering
\small
\setlength{\tabcolsep}{6pt}
\renewcommand{\arraystretch}{1.08}
\caption{Reporting sufficiency for clean cross-paper latency comparison. The rule requires a headline claim, baseline, latency, communication, end-to-end rounds, network context, cost scope, and amortization context. This table measures sufficiency for comparison, not paper quality.}
\label{tab:reporting-sufficiency}
\begin{tabular}{lcc}
\toprule
\textbf{Status} & \textbf{Papers} & \textbf{Share} \\
\midrule
Sufficient & 1/62 & 1.6\% \\
Partial & 43/62 & 69.4\% \\
Insufficient & 18/62 & 29.0\% \\
\bottomrule
\end{tabular}
\end{table}

\begin{table*}[t]
\centering
\small
\setlength{\tabcolsep}{5pt}
\renewcommand{\arraystretch}{1.08}
\caption{Sensitivity analysis for the two central conclusions: direct global leaderboard-style comparison is unsupported under published evidence, and progress claims are multi-causal. \emph{Protocol/mixed} counts papers whose attributions include protocol progress (as primary or secondary factor) or whose primary label is mixed; \emph{non-protocol primary} counts papers whose primary label is any non-protocol category. The reviewer-favorable row is an optimistic upper bound, not the locked result.}
\label{tab:sensitivity-analysis}
\resizebox{\textwidth}{!}{%
\begin{tabular}{lcccc}
\toprule
\textbf{Scenario} & \textbf{Competitive relations} & \textbf{Baseline-compatible claims} & \textbf{Protocol/mixed} & \textbf{Non-protocol primary} \\
\midrule
Locked labels & 1/1891 & 0/62 & 52/62 & 40/62 \\
Optimistic progress attribution & 1/1891 & 0/62 & 52/62 & 40/62 \\
Strict progress attribution & 1/1891 & 0/62 & 52/62 & 40/62 \\
Optimistic comparability & 93/1891 & 2/62 & 52/62 & 40/62 \\
Strict comparability & 1/1891 & 0/62 & 52/62 & 40/62 \\
Drop moderate/boundary confidence & 1/1891 & 0/62 & 45/49 & 31/49 \\
Reviewer-favorable upper bound & 93/1891 & 2/62 & 52/62 & 40/62 \\
\bottomrule
\end{tabular}
}
\end{table*}

\subsection{PIC-Card Field Distributions}
\label{app:pic_distributions}
\begin{figure}[!b]
\centering
\includegraphics[width=\columnwidth]{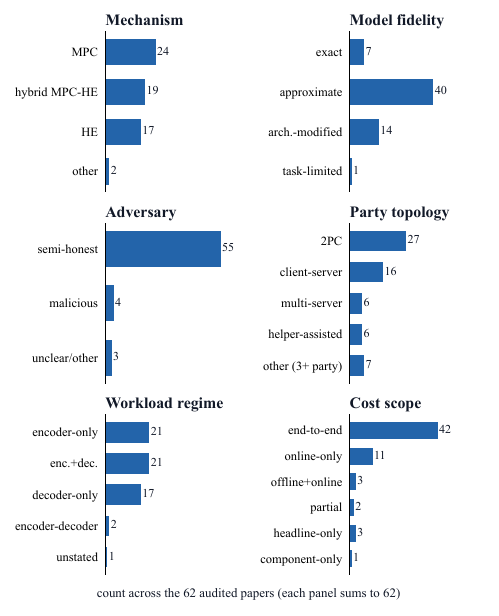}
\caption{\textbf{Distribution of the 62 audited papers over the categorical
PIC-card fields.} The corpus is near-constant on the threat axes that rarely block
comparison (55/62 semi-honest; 49/62 protect both input and model) but spreads
across mechanism, model fidelity, party topology, and workload regime, the fields
that carry the comparability gates. Long tails are consolidated: ``other'' under
party topology aggregates five settings with three or more parties.}
\label{fig:pic-distributions}
\end{figure}

Figure~\ref{fig:pic-distributions} shows how the 62 audited papers distribute over
the categorical PIC-card fields, complementing the field statistics discussed in
Section~\ref{sec:results_graph_anatomy}.

\subsection{Full Comparability Matrix}
\label{app:comparability_matrix}
\begin{figure*}[t]
\centering
\includegraphics[width=\textwidth]{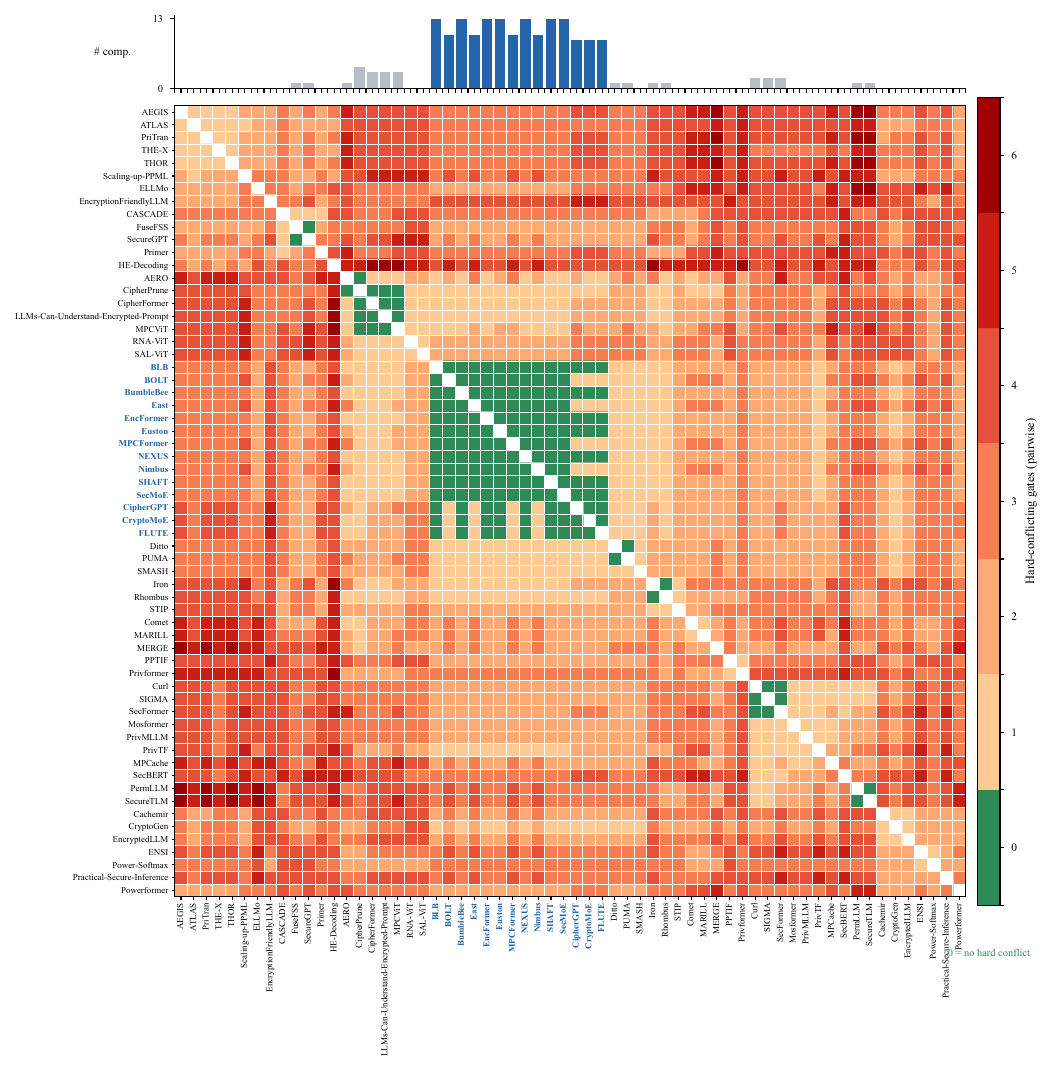}
\caption{\textbf{Full comparability graph with all 62 papers labeled.} As in
Figure~\ref{fig:comparability-heatmap}, each cell shades a pair by the number of
comparability gates that hard-conflict (green marks no hard conflict), with rows
and columns ordered by agglomerative clustering and the top bar counting each
paper's hard-conflict-free partners (hub systems in blue).}
\label{fig:comparability-heatmap-full}
\end{figure*}

Figure~\ref{fig:comparability-heatmap-full} is the fully labeled version of the
comparability graph summarized in Figure~\ref{fig:comparability-heatmap}, with all
62 papers named on both axes so individual pairings can be traced.

\subsection{Blind Re-Extraction Agreement}
\label{app:verification}
Table~\ref{tab:verification-agreement} reports the per-field agreement between the
recorded values and the independent blind re-extraction described in
Section~\ref{sec:method_validation}.
\begin{table*}[t]
\centering
\small
\setlength{\tabcolsep}{6pt}
\renewcommand{\arraystretch}{1.08}
\caption{Per-field agreement of the independent blind re-extraction with the
recorded human values, over the 62-paper corpus. \emph{Agree}/\emph{Disagree}
are value-vs-value; \emph{V.\ silent} is where the blind reader declined to
extract a value the human took from context (e.g., a parameter count implied by
a model name); \emph{Both silent} is concordant agreement that the paper states
nothing; \emph{H.\ silent} is where no human value was recorded (both excluded
from the rate). The rate is $\text{agree}/(\text{agree}+\text{disagree}+
\text{V.\ silent})$. The fields the comparability gates lean on most
(cryptographic backbone, privacy scope, adversary model, end-to-end status)
agree in 87--98\% of papers, with 70.0\% raw agreement
pooled across all factual fields; lower rates are dominated by values the blind
reader declines to infer and by concordant silence. Every flagged cell was
adjudicated against the source papers; across the two adjudication passes,
40 of 992 fields were corrected corpus-wide. The final row is the judgment-label inter-rater agreement, not a
factual field.}
\label{tab:verification-agreement}
\begin{tabular}{lcccccc}
\toprule
\textbf{Contract field} & \textbf{Agree} & \textbf{Disagree} & \textbf{V.\ silent} & \textbf{Both silent} & \textbf{H.\ silent} & \textbf{Rate} \\
\midrule
Cryptographic backbone & 61 & 1 & 0 & 0 & 0 & 98.4\% \\
Inference regime & 37 & 1 & 23 & 1 & 0 & 60.7\% \\
Privacy scope & 58 & 4 & 0 & 0 & 0 & 93.5\% \\
Adversary model & 54 & 2 & 6 & 0 & 0 & 87.1\% \\
Party model & 39 & 19 & 4 & 0 & 0 & 62.9\% \\
Model fidelity & 46 & 16 & 0 & 0 & 0 & 74.2\% \\
Cost scope & 38 & 20 & 4 & 0 & 0 & 61.3\% \\
Amortization & 12 & 7 & 0 & 43 & 0 & 63.2\% \\
Network regime & 26 & 9 & 7 & 20 & 0 & 61.9\% \\
Largest private model & 45 & 14 & 3 & 0 & 0 & 72.6\% \\
Largest model, params (B) & 22 & 8 & 32 & 0 & 0 & 35.5\% \\
End-to-end status & 54 & 8 & 0 & 0 & 0 & 87.1\% \\
Latency reported & 62 & 0 & 0 & 0 & 0 & 100.0\% \\
Communication reported & 42 & 20 & 0 & 0 & 0 & 67.7\% \\
Rounds reported & 10 & 52 & 0 & 0 & 0 & 16.1\% \\
Progress attribution (label) & 37 & 25 & 0 & 0 & 0 & 59.7\% \\
\bottomrule
\end{tabular}
\end{table*}

\begin{table*}[t]
\centering
\small
\renewcommand{\arraystretch}{1.2}
\caption{Prescriptive backbone guidance under the contract view
(Section~\ref{sec:discussion_backbone_choice}). These are cost-structure
tendencies conditioned on the deployment contract, not a universal ranking: the
preferable backbone is the one whose cost structure matches the dominant
constraint, which is also why a single speed leaderboard misleads.}
\label{tab:backbone-guidance}
\begin{tabularx}{\textwidth}{p{3.7cm} p{3.7cm} X}
\toprule
\textbf{Dominant deployment constraint} & \textbf{Tends to favor} & \textbf{Why} \\
\midrule
Interactive, WAN / high round-trip time & Low-round HE or few-round hybrid & Each interaction round pays the round-trip, so round-heavy MPC inflates wall-clock latency even when its LAN numbers look strong. \\
Batch / high throughput, amortizable & HE with SIMD packing; amortized-offline hybrid & Ciphertext batching and reusable preprocessing amortize across many queries, where per-query latency is secondary. \\
LAN / high bandwidth, low round-trip time & MPC or hybrid & Cheap rounds make interactive protocols competitive or fastest; HE ciphertext and compute overhead rarely pays off here. \\
Bandwidth-constrained link & HE or low-communication designs & MPC communication volume can dominate cost; HE trades bytes for local computation. \\
Model weights must stay private & Input-and-model 2PC (HE or hybrid); avoid weight-revealing split or statistical schemes & Protecting weights is a hard constraint largely orthogonal to speed; some fast schemes assume public weights. \\
A non-colluding helper or third party is acceptable & 3PC or helper-assisted MPC & A helper trusted for non-collusion unlocks cheaper nonlinear and truncation protocols. \\
\bottomrule
\end{tabularx}
\end{table*}

\subsection{PTI frontiers over the year}
\label{sec:app-frontier-models}
Table~\ref{tab:scaling-contracts} summarizes the largest PTI systems (by parameter count) from 2022 through 2024. As shown, these milestone models differ substantially across their underlying cryptographic protocols, party topologies, model fidelity, and cost scopes.
\begin{table*}[t]
\centering
\small
\setlength{\tabcolsep}{6pt}
\renewcommand{\arraystretch}{1.1}
\caption{The conventional cryptographic PTI frontier of
Figure~\ref{fig:scaling-linear}, annotated with each frontier point's evaluation
contract. The points span different party models (2PC, 3PC, helper-assisted) and
cost scopes (end-to-end, online-only), so the cumulative envelope is a generous
upper bound over heterogeneous contracts, not a single comparable trajectory.
Tightening to one fixed contract could only lower the private-inference frontier.}
\label{tab:scaling-contracts}
\begin{tabular}{llrllll}
\toprule
\textbf{Year} & \textbf{System} & \textbf{Params} & \textbf{Mechanism} & \textbf{Party model} & \textbf{Model fidelity} & \textbf{Cost scope} \\
\midrule
2022 & Iron  & 0.34B & Hybrid MPC--HE & 2PC             & exact       & end-to-end  \\
2023 & PUMA  & 7B    & MPC            & 3PC             & approximate & end-to-end  \\
2024 & SIGMA & 13B   & MPC            & helper-assisted & approximate & online-only \\
\bottomrule
\end{tabular}
\end{table*}

\subsection{Robustness of the Findings}
\label{sec:results_sensitivity}
The three findings above rest on labels we assigned by hand, so we close by
recomputing the central numbers under different labeling assumptions
(Table~\ref{tab:sensitivity-analysis}, Appendix~\ref{app:extra_tables}). The
locked setting gives 1/1891 competitive relations and 0/62 baseline-compatible
primary claims, and the strict setting matches it. The optimistic setting waves
every unknown gate through, and even then only 93/1891 relations turn competitive
and only 2/62 claims become baseline-compatible. The reviewer-favorable upper
bound, which stacks optimistic attribution, comparability, and baseline
compatibility together, lands on the same 93/1891 and 2/62, with non-protocol
primary factors still present in 40/62 papers. Drop the moderate- and
boundary-confidence rows and 49 papers remain, telling the same story: 1/1891
competitive, 45/49 protocol-or-mixed, 31/49 non-protocol-primary. The conclusions
do not ride on a single pessimistic call.

\smallskip\noindent
Put the three findings together and they say one thing: \emph{PTI progress is real,
but the published evidence does not yet show cumulative progress toward
deployment-scale private LLM inference}. The frontier sits far below public LLM
scale, comparability breaks at specific contract fields, the gains come from many
sources at once, and none of it hinges on fragile labels. The next section steps
past the published record: Section~\ref{sec:cost_roadmap} puts the scaling gap in
engineering units.

\section{Reliability of the Cost Projections}
\label{app:cost-validation}
The production-scale figures in Section~\ref{sec:cost_roadmap} are projections
rather than measurements, so two checks tell us how far to trust them. First, the
power-law form holds \emph{within} the measured range: refit the SIGMA size series
on its three smaller models (0.124--7B), predict the held-out 13B point, and you
land within 7.7\% on time, 2.5\% on communication, and 5.3\% on key size, with
leave-one-out error never above 12\%. Second, the qualitative gap survives whatever
exponent we pick for the single-anchor rows. Sweep the structural exponent from
sub-linear ($0.8$) to super-linear ($1.2$) and a projected 70B forward pass only
moves between tens of minutes (3PC) and a few hours (hybrid 2PC), and a 405B pass
between about an hour and a day. The multiplier shifts; the order of magnitude does
not. They are still extrapolations, of course: 70B and 405B sit $5\times$ and
$31\times$ past the largest model we measured, and no power law can see a regime
change coming, a bootstrapping threshold crossed, a packing or ring-parameter
switch, a memory ceiling. So we read the result as a statement about \emph{scale},
three to five orders of magnitude away from plaintext, rather than as a specific
predicted runtime.

\section{Field Extraction Protocol and Codebook}
\label{app:llm_protocol}

This appendix documents the structured field-extraction protocol behind the audit,
included verbatim for reproducibility. The codebook below is the instrument the
extractor works from: it lists the fields we attempt for each paper, organized into
priority tiers, and the prompt spells out how each field is read off the source
text. Extraction is evidence-first throughout: every recorded value is tied to an
explicit statement in the source paper, and anything missing is recorded as missing
rather than guessed at. Coverage therefore varies by field. The Tier-0 contract
fields are recorded for every paper and carry the comparability, attribution, and
scaling analyses; the quantitative performance tuples feed the cost model on its
calibration subset (Section~\ref{sec:cost_roadmap}). Many Tier-1 and Tier-2 fields
are only sporadically reported in the source papers, so they are frequently marked
NOT\_REPORTED and are not used in this paper's results.

\subsection{Quantitative Extraction Prompt}

\begin{promptbox}
You are extracting quantitative performance metrics for a privacy-preserving Transformer inference paper.

You will be provided:
(1) The paper text
(2) the Codebook -- Quantitative Performance Extraction (Raw Metrics)

CRITICAL INSTRUCTIONS -- READ CAREFULLY:

1) Output ONLY ONE valid JSON object.
   - No markdown, no backticks
   - No explanations, no commentary
   - No text before or after JSON

2) Use ONLY explicit statements from the paper (and the provided Codebook).
   - Never infer, never estimate, never approximate
   - Never compute unless the Codebook explicitly allows it

3) Missing tokens (case-sensitive): use EXACTLY one when needed:
   NOT_REPORTED
   NOT_SEPARATED
   NOT_POSSIBLE
   THROUGHPUT_ONLY
   PER_LAYER_ONLY
   QUALITATIVE_ONLY
   UNSPECIFIED

4) All numeric values must be numbers (not strings) unless using missing tokens.

5) Evidence requirements:
   a) For every non-missing RAW numeric or boolean field, you MUST include an entry in the "evidence" array.
   b) Each evidence entry MUST contain:
      - field_path (must match the JSON hierarchy EXACTLY)
      - evidence_quote (<=25 words)
      - evidence_location (Table X / Fig Y / Sec Z / p.N)
   c) If a RAW numeric/boolean field lacks evidence, mark that field as NOT_REPORTED (or the correct missing token).

5.1) FIELD_PATH STRICTNESS (MANDATORY):

   - The "field_path" value in each evidence entry MUST exactly match the schema path.
   - Use dot notation that mirrors the JSON structure, for example:
       "Q0.1.model_name"
       "Q0.3.e2e_latency_value"
       "Q1.5.num_rounds"
       "Q2.3.code_available"
   - Do NOT prefix with an extra level (e.g., NOT "Q0.Q0.1.model_name").
   - Do NOT invent new paths.
   - The field_path must correspond exactly to an existing leaf field in the provided schema.
   - If you cannot match a field_path exactly, mark the field as NOT_REPORTED.

6) Normalized fields policy:
   - normalized_* fields may be filled ONLY if unit conversion is unambiguous.
   - Evidence is required for the RAW reported value.
   - A separate evidence entry for the normalized field is NOT required if the raw field is properly evidenced.

7) Choose EXACTLY ONE primary evaluation setting using the Codebook's selection rule.
   Provide:
   - setting_label (short identifier, no spaces)
   - chosen_setting_rationale (1-2 lines referencing the selection rule)

8) Strict completeness rule:
   - ALL fields in the schema MUST be present.
   - Do NOT remove or add fields.
   - Do NOT reorder structure.
   - Use null only where explicitly allowed in the schema.
   - Never leave a field empty.

9) If you cannot produce valid JSON, output EXACTLY:
   {"error":"INVALID_JSON"}

Now extract according to the schema below.
Return ALL fields exactly as structured.

{
  "paper_id": "<FILL>",
  "setting_label": "<SHORT_LABEL_FOR_SELECTED_SETTING>",
  "chosen_setting_rationale": "<1-2 lines>",

  "Q0": {
    "Q0.1": {
      "model_name": "UNSPECIFIED",
      "num_parameters": "NOT_REPORTED",
      "model_type": "UNSPECIFIED"
    },
    "Q0.2": {
      "task_type": "UNSPECIFIED",
      "sequence_length": "NOT_REPORTED",
      "batch_size": "NOT_REPORTED",
      "num_generated_tokens": "NOT_REPORTED",
      "notes_on_setting": null
    },
    "Q0.3": {
      "e2e_latency_value": "NOT_REPORTED",
      "e2e_latency_unit": "NOT_REPORTED",
      "e2e_latency_granularity": "UNSPECIFIED",
      "includes_offline": "NOT_REPORTED",
      "normalized_e2e_latency_ms": "NOT_POSSIBLE"
    },
    "Q0.4": {
      "total_comm_value": "NOT_REPORTED",
      "total_comm_unit": "NOT_REPORTED",
      "comm_direction": "NOT_REPORTED",
      "normalized_total_comm_MB": "NOT_POSSIBLE"
    },
    "Q0.5": {
      "network_type": "NotSpecified",
      "bandwidth_value": "NOT_REPORTED",
      "bandwidth_unit": "NOT_REPORTED",
      "rtt_latency_ms": "NOT_REPORTED"
    },
    "Q0.6": {
      "server_hardware": "NOT_REPORTED",
      "client_hardware": "NOT_REPORTED",
      "accelerator": "NotSpecified",
      "deployment": "NotSpecified"
    },
    "Q0.x": {
      "model_dependency": "NOT_REPORTED",
      "phase_scope": "NOT_REPORTED",
      "multi_client_scaling_evaluated": "NOT_REPORTED",
      "comparability_limits_noted": "NOT_REPORTED",
      "client_privacy_explicit": "NOT_REPORTED",
      "server_privacy_explicit": "NOT_REPORTED"
    }
  },

  "Q1": {
    "Q1.1": {
      "online_latency_value": "NOT_SEPARATED",
      "online_latency_unit": "NOT_REPORTED",
      "online_latency_granularity": "UNSPECIFIED"
    },
    "Q1.2": {
      "offline_latency_value": "NOT_REPORTED",
      "offline_latency_unit": "NOT_REPORTED",
      "offline_amortized": "NOT_REPORTED"
    },
    "Q1.3": {
      "server_compute_time_value": "NOT_SEPARATED",
      "server_compute_time_unit": "NOT_REPORTED"
    },
    "Q1.4": {
      "client_compute_time_value": "NOT_SEPARATED",
      "client_compute_time_unit": "NOT_REPORTED"
    },
    "Q1.5": {
      "num_rounds": "NOT_REPORTED",
      "num_messages": "NOT_REPORTED"
    },
    "Q1.6": {
      "peak_server_memory_value": "NOT_REPORTED",
      "peak_server_memory_unit": "NOT_REPORTED",
      "peak_client_memory_value": "NOT_REPORTED",
      "peak_client_memory_unit": "NOT_REPORTED"
    },
    "Q1.7": {
      "throughput_value": "NOT_REPORTED",
      "throughput_unit": "NOT_REPORTED",
      "throughput_setting": null
    },
    "Q1.8": {
      "utility_metric_name": "NOT_REPORTED",
      "utility_value": "NOT_REPORTED",
      "utility_unit": "NOT_REPORTED",
      "utility_task_or_dataset": "NOT_REPORTED",
      "delta_vs_plaintext_value": "NOT_REPORTED",
      "delta_unit": "NOT_REPORTED"
    }
  },

  "Q2": {
    "Q2.1": {
      "num_bootstraps": "NOT_REPORTED",
      "multiplicative_depth": "NOT_REPORTED",
      "num_ciphertext_mults": "NOT_REPORTED",
      "ciphertext_size_bytes": "NOT_REPORTED"
    },
    "Q2.2": {
      "runtime_cost_usd": "NOT_REPORTED",
      "pricing_context": "NOT_REPORTED"
    },
    "Q2.3": {
      "code_available": "unknown",
      "artifact_link": "NOT_REPORTED",
      "runnable_claim": "unknown"
    }
  },

  "extraction_warnings": [],
  "evidence": []
}

\end{promptbox}

\subsection{Quantitative Performance Extraction Codebook}

\begin{promptbox}
the Codebook -- Quantitative Performance Extraction (Raw Metrics)
Scope: Transformer private inference systems (end-to-end or near end-to-end).
Strict rule: Use only explicitly stated numbers/claims. Never infer. If missing -> NOT_REPORTED.
Global extraction rules
1.	Primary setting selection rule (when multiple settings exist):
o	Prefer end-to-end over microbenchmarks.
o	Prefer largest model evaluated.
o	Prefer WAN setting over LAN if both are given.
o	Prefer main headline benchmark used in the paper's abstract/intro/eval summary.
2.	Units: Store as reported in raw fields + also store a normalized_* field when conversion is unambiguous (e.g., s->ms, GB->MB).
3.	Evidence required: every non-missing (i.e., not NOT_REPORTED/NOT_SEPARATED/etc.) numeric/boolean field must include evidence_quote and evidence_location.
4.	If model name implies standard parameter count (e.g., ``BERT-base'', ``LLaMA-7B''), you may fill num_parameters using the canonical model size only if the paper explicitly uses that model name, and include evidence pointing to the model name.
________________________________________
Tier-0 (must attempt)
Q0.1 -- Model Architecture
Extract fields
-	model_name (string | UNSPECIFIED)
-	num_parameters (integer | NOT_REPORTED)
-	model_type in {Encoder, Decoder, Encoder-Decoder, VisionTransformer, Other, UNSPECIFIED}
Q0.2 -- Workload / Input Configuration
-	task_type in {NLP_classification, LLM_generation, Vision, Other, UNSPECIFIED}
-	sequence_length (int | NOT_REPORTED)
-	batch_size (int | NOT_REPORTED)
-	num_generated_tokens (int | NOT_REPORTED) only if generation
-	notes_on_setting (short string | null)
Q0.3 -- End-to-End Latency (primary)
Definition: total time per inference query/batch including crypto operations.
-	e2e_latency_value (number | THROUGHPUT_ONLY | NOT_REPORTED)
-	e2e_latency_unit in {ms, s, us, min, NOT_REPORTED}
-	e2e_latency_granularity in {per_query, per_batch, per_token, UNSPECIFIED}
-	includes_offline in {yes, no, unclear, NOT_REPORTED}
-	normalized_e2e_latency_ms (number | NOT_POSSIBLE)
Q0.4 -- Total Communication (end-to-end)
-	total_comm_value (number | PER_LAYER_ONLY | NOT_REPORTED)
-	total_comm_unit in {B, KB, MB, GB, NOT_REPORTED}
-	comm_direction in {bidirectional_total, client_to_server_only, server_to_client_only, unknown, NOT_REPORTED}
-	normalized_total_comm_MB (number | NOT_POSSIBLE)
Q0.5 -- Network Setting
-	network_type in {LAN, WAN, Simulated, NotSpecified}
-	bandwidth_value (number | NOT_REPORTED)
-	bandwidth_unit in {Mbps, Gbps, NOT_REPORTED}
-	rtt_latency_ms (number | NOT_REPORTED)
Q0.6 -- Hardware Configuration
-	server_hardware (string | NOT_REPORTED)
-	client_hardware (string | NOT_REPORTED)
-	accelerator in {CPU, GPU, TPU, Mixed, NotSpecified}
-	deployment in {Cloud, Local, NotSpecified}
________________________________________
Tier-1 (high-value diagnostics; attempt if available)
Q1.1 -- Online Latency
-	online_latency_value, online_latency_unit, online_latency_granularity
-	If not separated -> NOT_SEPARATED
Q1.2 -- Offline / Preprocessing Latency
-	offline_latency_value, offline_latency_unit
-	offline_amortized in {yes, no, unclear, NOT_REPORTED}
Q1.3 -- Server Compute Time
-	server_compute_time_value, unit
-	If not separated -> NOT_SEPARATED
Q1.4 -- Client Compute Time
-	client_compute_time_value, unit
-	If not separated -> NOT_SEPARATED
Q1.5 -- Interaction Depth
-	num_rounds (int | NOT_REPORTED)
-	num_messages (int | NOT_REPORTED)
Q1.6 -- Peak Memory Usage
-	peak_server_memory_value, unit in {MB, GB, NOT_REPORTED}
-	peak_client_memory_value, unit in {MB, GB, NOT_REPORTED}
Q1.7 -- Throughput (if reported)
-	throughput_value (number | NOT_REPORTED)
-	throughput_unit in {queries_per_s, tokens_per_s, seq_per_s, img_per_s, NOT_REPORTED}
-	throughput_setting (string | null)
Q1.8 -- Utility / Accuracy (if evaluated)
-	utility_metric_name (string | NOT_REPORTED)
-	utility_value (number | NOT_REPORTED)
-	utility_unit (string | NOT_REPORTED) (e.g., %, ppl)
-	utility_task_or_dataset (string | NOT_REPORTED)
-	delta_vs_plaintext_value (number | NOT_REPORTED)
-	delta_unit in {pp, absolute, relative_percent, NOT_REPORTED}
________________________________________
Tier-2 (optional / sparse / crypto-primitive-level)
Q2.1 -- Crypto primitive counts / depths (only if explicitly numeric)
-	num_bootstraps, multiplicative_depth, num_ciphertext_mults, ciphertext_size_bytes
-	If described qualitatively only -> QUALITATIVE_ONLY
Q2.2 -- Monetary cost (only if explicit)
-	runtime_cost_usd (number | NOT_REPORTED)
-	pricing_context (string | NOT_REPORTED)
Q2.3 -- Reproducibility metadata (quantifiable)
-	code_available in {yes, no, partial, unknown}
-	artifact_link (string | NOT_REPORTED)
-	runnable_claim in {docker, scripts, not_specified, unknown}
________________________________________
Standard output JSON skeleton (one paper)
{
  "paper_id": "STRING",
  "chosen_setting_rationale": "1-2 lines citing the selection rule used",
  "Q0": { /* includes Q0.1-Q0.6 and Q0.x */ },
  "Q1": { },
  "Q2": { },
  "extraction_warnings": ["..."],
  "evidence": [
    {
      "field_path": "Q0.3.e2e_latency_value",
      "evidence_quote": "<=25 words",
      "evidence_location": "Table X / Fig Y / Sec Z / p.N"
    }
  ]
}


Q0.x -- Technical synthesis tags (Pass A)
These fields are added to support technical (not just clarity) findings and cross-paper synthesis. They are intentionally simple and evidence-first: if not explicitly stated, mark NOT_REPORTED.

New categorical fields (Pass A)
- model_dependency: {MODEL_SPECIFIC, ARCHITECTURE_SPECIFIC, MODEL_AGNOSTIC, NOT_REPORTED}
- phase_scope: {INFERENCE_ONLY, TRAINING_ONLY, BOTH, NOT_REPORTED}
- multi_client_scaling_evaluated: {yes, no, NOT_REPORTED}
- comparability_limits_noted: short free-text (1 sentence) or NOT_REPORTED
- client_privacy_explicit: {yes, no, unclear, NOT_REPORTED}
- server_privacy_explicit: {yes, no, unclear, NOT_REPORTED}

Notes: (i) MODEL_AGNOSTIC means the protocol is described as applicable to multiple model families without retraining or per-architecture retooling; (ii) ARCHITECTURE_SPECIFIC means it targets a class (e.g., Transformers, CNNs) but not a single named model; (iii) MODEL_SPECIFIC means the technique/evaluation is tied to a specific model instance (e.g., BERT-base) or requires model-specific engineering.

\end{promptbox}


\section{Audit Case Studies}
\label{sec:audit_case_studies}

This section walks through four comparison patterns that turn up again and again
in the corpus, and shows how the gates change the reading of each. The patterns
are schematic on purpose; they are not meant to call out individual papers. Each
one starts from a comparison that looks perfectly natural at the headline level,
and then shows what the restored contract has to say about it.
Table~\ref{tab:audit-patterns} gathers all four.

\begin{table*}[t]
\centering
\small
\setlength{\tabcolsep}{4pt}
\renewcommand{\arraystretch}{1.15}
\caption{Four recurring comparison patterns and their contract reading. Each
looks like a direct comparison at the headline level; the gates identify the
mismatched or missing field that makes it descriptive or invalid.}
\label{tab:audit-patterns}
\begin{tabularx}{\textwidth}{p{2.6cm} X X p{1.6cm} X}
\toprule
\textbf{Pattern} & \textbf{Why it looks comparable} & \textbf{Contract issue} & \textbf{Gate} & \textbf{Sound reading} \\
\midrule
Online latency vs.\ end-to-end cost &
Both report private-inference latency. &
One number is online-only or amortized; the other counts more of the end-to-end
cost or assumes different reuse. &
G5, G6 &
Not directly comparable unless cost scope and reuse assumptions are aligned. \\
\midrule
LAN result vs.\ WAN claim &
Both report latency for the same broad model family. &
The measurement is LAN, simulated, or unspecified while the motivating
deployment is WAN-like; rounds convert round-trip time into wall-clock cost. &
G7, G8 &
Supports a LAN or local-deployment claim, not an unrestricted WAN claim. \\
\midrule
Encoder classification vs.\ decoder generation &
Both are Transformer systems built on secure-computation primitives. &
The workloads differ in regime and in sequence/generation semantics
(per-token vs.\ per-sequence vs.\ per-batch). &
G3 (often G4) &
Different comparison neighborhoods unless workload and fidelity are normalized. \\
\midrule
Model-preserving vs.\ model-modifying &
Both target the same model family or benchmark. &
The faster result depends on approximation, retraining, pruning, distillation,
quantization, or architecture change. &
G4 &
Possibly workload/fidelity-shift progress; not pure protocol progress. \\
\bottomrule
\end{tabularx}
\end{table*}

Two of these patterns carry a wider lesson. The encoder-versus-decoder pattern is
why the taxonomy in Table~\ref{tab:taxonomy-ppti} cannot settle comparability on its
own: the taxonomy tells you two systems live in the same design space, while the
gates tell you whether their claims answer the same question. The fidelity pattern
is the bridge from comparability to attribution. Modifying a model may well be the
right way to make private inference practical, but the paper has to name that gain
as fidelity-shift progress. Otherwise a reader chalks the whole speedup up to
cryptographic protocol progress, and the literature ends up recording a protocol
advance the evidence never showed.

In every one of these patterns, the failure is not that a paper measured the wrong
thing. It happens at comparison time, the moment a reader folds two numbers
measured under different contracts into a single ranking. What the gates do is make
that step visible, and auditable.

\end{document}